\documentclass[showpacs,pre,amssymb,amsmath,preprint]{revtex4-1}
\usepackage{amsmath}
\usepackage{amssymb}
\usepackage{array}		
\usepackage{multirow}		
\usepackage{rotating}		
\usepackage{color}
\usepackage{Mayer}
\usepackage{cases}

\newcommand{\be}{\begin{equation}}
\newcommand{\ee}{\end{equation}}
\newcommand{\bea}{\begin{eqnarray}}
\newcommand{\eea}{\end{eqnarray}}
\def \la{\label}

\def\({\left (}
\def\){\right )}
\def\]{\right]}
\def\[{\left[}
\def\<{\left <}
\def\>{\right>}

\newcommand{\bx}{\mathbf{x}}

\newcommand{\br}{\mathbf{r}}

\newcommand{\bk}{\mathbf{k}}

\newcommand{\cE}{\mathcal{E}}

\newcommand{\cS}{\mathcal{S}}

\newcommand{\cL}{\mathcal{L}}

\newcommand{\bX}{\mathbf{X}}
\newcommand{\bcX}{\pmb{\mathcal{X}}}

\renewcommand{\d}{\mathrm{d}}

\newcommand{\Tr}{\mathrm{Tr}}
\newcommand{\e}{\mathrm{e}}

\newcommand{\sx}{\textsf{\textbf{x}}}

\makeatletter
\newcommand{\xRightarrow}[2][]{\ext@arrow 0359\Rightarrowfill@{#1}{#2}}
\makeatother

\begin{document}

  \title{Screened activity expansion for the grand-potential of a quantum plasma and how to derive approximate
	equations of state compatible with electroneutrality}

\author{A. Alastuey$^{1}$, V. Ballenegger$^2$ and D. Wendland$^{1,2}$ 
       \\[2mm]
       $^1${\small Laboratoire de Physique, ENS Lyon, UMR CNRS 5672}\\[-1mm]
        {\small 46 all\'ee d'Italie, 69364 Lyon Cedex 07, France}\\[-1mm]
              $^2${\small Institut UTINAM, Univ. Bourgogne-Franche-Comt\'e, UMR CNRS 6213} \\[-1mm] 
							{\small 16, route de Gray, 25030 Besan\c{c}on Cedex, France}}

  \date{\today}
\begin{abstract}

We consider a quantum multi-component plasma made with $\cS$ species of point charged particles interacting \textit{via} the 
Coulomb potential.  We derive the screened activity series for the pressure in the grand-canonical ensemble 
within the Feynman-Kac path integral representation of the system in terms of a classical gas of loops. 
This series is useful for computing equations of state for it is non-perturbative with respect to the strength of the interaction and it involves relatively few diagrams at a given order. The known screened activity series for the particle densities can be recovered by differentiation. The particle densities satisfy local charge neutrality thanks to a Debye-screening dressing mechanism of the diagrams in these series. We introduce a new general neutralization prescription, based on this mechanism, for deriving approximate equations of state where consistency with electroneutrality is automatically ensured. This prescription is compared to other ones, including a neutralization scheme inspired by the Lieb-Lebowitz theorem and based on the introduction of $(\cS-1)$ suitable independent combinations of the activities. Eventually, we briefly argue how the activity series for the pressure, combined with the Debye-dressing prescription, can be used for deriving approximate equations of state at moderate densities, which include the contributions of recombined entities made with three or more particles. 

\end{abstract}
\maketitle



\section{Introduction}

The thermodynamic properties of hydrogen and helium gases enter as a basic and key ingredient in the study of the structure and evolution of dense astrophysical bodies like stars, brown dwarfs and giant planets, for these objects are mostly made of a mixture of H and He. Many works have been performed to determine the equation of state of H-He mixtures and also to address the related question of the helium solubility in hydrogen, which is important for a correct description of planetary interiors~\cite{Fantoni2015,Morales2013,Lorenzen2009}. The hydrogen-helium mixture is fundamentally a quantum plasma made of electrons and nuclei interacting via the Coulomb potential. Numerical simulation techniques, like density-functional theory molecular dynamics~\cite{Militzer2013,Lorenzen2009,Vorberger2007,Lenosky2000,Perrot1995}, Path Integral Monte Carlo~\cite{Militzer2001,Militzer2000b,Pierleoni1994} and Quantum Monte Carlo~\cite{Mazzola2018}, have been used to calculate, with good precision, some thermodynamical properties of H-He mixtures in strongly interacting regimes.  Besides simulations, analytical calculations, in particular asymptotic expansions, are useful to provide theoretical insights and to complement the simulation data with reliable results in asymptotic regimes, like at low density or at high density or high temperature~\cite{Alastuey2012,Chabrier2019}. Such expansions have been derived using various analytical tools: the effective potential method~\cite{Morita1959,Ebeling1967,Deutsch1977}, many-body perturbation theory~\cite{deWitt1995,Kremp2005,Alastuey1994} and two different path-integral formalisms: Mayer diagrammatical expansions in the ring-polymer representation~\cite{Alastuey1994,Alastuey1996,Brydges1999,Ballenegger2002} and an effective field theory~\cite{Brown2001}.  It has been checked explicitly that these quite different theoretical frameworks all lead to the same expansion at low densities~\cite{Alastuey2015}. Beyond asymptotic expansions at low or high densities, analytical theories can also provide insights at intermediate densities via the introduction of suitable approximations (see e.g. \cite{Ebeling2012,Alastuey2012a,Ramazanov2015,Bunker1997}).

\bigskip

In this article, we derive two general results that enable easier (full or partially) analytical calculations of the equation of state of H-He mixtures, and also other plasmas, in the low and moderate density regimes. First, we obtain a new exact representation for all terms in the activity expansion of the grand-potential $\Omega = -P \Lambda$ ($\Lambda$ denotes the volume) of a quantum multi-component plasma. This so-called screened Mayer activity-series for $\Omega$, or equivalently for the pressure $P$, complements  the activity-series for distribution functions derived in Ref.~\cite{Ballenegger2002} and provides a much more direct route for computing the equation of state: less diagrams need to be computed and no term-by-term integration of contributions to distribution functions needs to be performed. Since $\Omega$ is a thermodynamic potential, all thermodynamic properties can furthermore be deduced from it via standard thermodynamic relations. The screened Mayer series are not perturbative with respect to the strength of the interaction, in contrast to the expressions of standard many-body perturbation theory (thermodynamic Green function formalism~\cite{Kremp2005}), and allow therefore calculations not only in the fully ionized regimes, at low or high densities, but also in moderately dense regimes where the particles are bound into atoms and/or molecules.

\bigskip

In the grand-canonical ensemble, the particle densities are deduced from the pressure thanks to the standard thermodynamical relation 
$\rho_\alpha = {\partial P}/{\partial \mu_\alpha}$. 
If charge neutrality is automatically satisfied for the exact expression of $P$, it is not necessarily the case for an approximate expression $P_{\rm A}$ of $P$. We introduce therefore, and this constitutes our second main result, general procedures for making any approximation $P_{\rm A}$ automatically consistent with electroneutrality. In these procedures, either dressed or neutral-group activities, are introduced based on general properties of quantum plasmas at equilibrium. 
We then deduce directly from $P_{\rm A}$, without any equation to solve, an associated thermodynamical potential that is compatible with electroneutrality.
We show that the various neutralization prescriptions do not lead in general to identical results for the equation of state.\footnote{unless the original approximation $P_{\rm A}$ is already compatible with electroneutrality, as for example in an asymptotic expansion, in which case these procedures are without effect.} The choice of a particular prescription is hence worthy of attention since it is not inconsequential.

\bigskip

The paper is organized as follows. In Section~\ref{sec:S2}, we define the model and recall that electroneutrality always holds in the bulk of a plasma, whatever the activities $\{z_\alpha \}$ are. This implies that one can impose the pseudo-neutrality condition $\sum_{\alpha=1}^{\cS} \e_\alpha z_\alpha = 0$ with $\cS$ the number of species, without loss of generality and that the average bulk properties necessarily depend only on
the temperature and on $(\cS-1)$ independent variables $\{y_i \}$, called neutral-group activities. An approximate expression~$P_{\rm A}(T;\{ z_{\alpha} \})$ of the pressure, which is not necessarily compatible with electroneutrality, can then be made compatible by adjusting it so that it depends on the activities only through 
$(\cS-1)$ neutral-group activities.
This defines the neutral-group neutralization prescription, which is new to our knowledge. This prescription is not unique for a plasma with three or more components because there are several ways,  when $\cS \geq 3$, of grouping particles together such that each group is charge neutral. 

\bigskip

The screened activity series for $P(T;\{ z_{\alpha} \})$ is derived in Section~\ref{sec:S3}. This series is obtained within the path-integral representation 
of the quantum system in terms of an equivalent classical gas of loops. This allows one to apply two standard classical techniques, namely Mayer diagrammatical expansions and Abe-Meeron summations~\cite{Mayer1940,Abe1959,Meeron1958}.
%
The known screened diagrammatic series for the particle densities~\cite{Alastuey2003} are recovered, as it should, 
by differentiating the present series for the pressure. The 
$z$-series for the particle densities do satisfy the local charge neutrality order by order, thanks to 
the combination of the pseudo-neutrality condition with a Debye-dressing mechanism. We
show how the well-known virial expansion of the EOS up to order $\rho^2$ can be recovered by 
keeping a few simple diagrams. 

\bigskip

In Section~\ref{DDR}, we introduce a Debye-Dressing prescription which automatically ensures that the particle densities 
inferred from any approximate function~$P_{\rm A}(T;\{ z_{\alpha} \})$ do satisfy local charge neutrality. This prescription is directly inspired by 
the Debye-dressing mechanism at work in the screened Mayer diagrammatic series. This method is compared to other 
procedures for ensuring electroneutrality, like the Neutral-Group procedure (Section~\ref{sec:S2}) and 
the Enforced-Neutrality method~\cite{Starostin2005}. 

\bigskip

In Section~\ref{Sec:S5}, we show how approximate equations of state at moderate densities 
can be constructed  by using the diagrammatic series 
for $P(T;\{ z_{\alpha} \})$, together with densities deduced via simple derivatives in which either the neutral-group activities (Section~II) or the Debye-Dressed activities (Section~IV) are used to ensure that electroneutrality is satisfied.
We point out that important physical mechanisms, like the recombination of nuclei and electrons into chemical species and atom-charge interactions, can be taken into account by retaining a few selected diagrams, in the spirit of the ACTEX method introduced by Rogers~\cite{Rogers1974,Rogers1981,Rogers1997,Rogers2002b} which underlie the OPAL thermodynamic tables~\cite{OPAL1996,Rogers2002}.
Some conclusions and perspectives are eventually given in Section~\ref{sec:S6}.

\section{Neutrality in the grand-canonical ensemble}\label{sec:S2}
\subsection{Quantum multi-component Coulomb system and the thermodynamic limit}

We consider a quantum multi-component plasma made of $\cS$ species of charged point particles enclosed 
in a box with volume $\Lambda$.   
The species index is denoted by $\alpha \in \{1,...,\cS\}$. Each particle of species $\alpha$ 
has a mass $m_\alpha$, while it carries a charge $e_\alpha$ and a spin $s_\alpha$. Each of them 
obeys to either Bose or Fermi statistics, according to the integer or half-integer value of $s_\alpha$ respectively. 
In order to ensure thermodynamic stability, at least one species needs to be fermions~\cite{Lieb1972} and there must be both positively and negatively
charged species. The species
$\alpha$ and the position $\bx$ of a given particle is denoted by the single notation 
$\sx=(\alpha,\bx)$. The total interaction potential $U(\sx_1,...,\sx_N)$ of $N$ particles 
is the sum of pairwise pure Coulomb interactions,
\be
\la{IX.QMG1}
U(\sx_1,...,\sx_N)= \sum_{i <j} V_{\rm C}(\sx_i,\sx_j)
\ee
with 
\be
\la{IX.QMG2}
V_{\rm C}(\sx_i,\sx_j)=e_{\alpha_i} e_{\alpha_j} v_{\rm C}(|\bx_i - \bx_j|)
\ee
and $v_{\rm C}(r)=1/r$. The corresponding non-relativistic Coulomb Hamiltonian reads
\be
H_{N}=-\sum_{i=1}^{N}\frac{\hbar^2}{2m_{\alpha_i}} \Delta_i + U(\sx_1,...,\sx_N)
\la{IX.QMG3}
\ee
where $\Delta_i$ is the Laplacian with respect to position $\bx_i$. The nucleo-electronic plasma 
is an example of such multi-component system, where the negative point charge are electrons,
(species $\alpha=\cS$) while all positive point charges are nuclei (species $\alpha=1,...,\cS-1$).

\bigskip

As proved by Lieb and Lebowitz~\cite{Lieb1972}, the present quantum multi-component plasma has a well-behaved thermodynamic 
limit (TL), and all statistical ensembles become equivalent in this limit. In the grand-canonical 
ensemble the TL is defined by fixing the chemical potentials $\mu_\alpha$ of each species as well as the 
inverse temperature $\beta=1/(k_{\rm B}T)$, and letting the volume $\Lambda \to \infty$. 
The grand-partition function $\Xi_{\Lambda}$ of the finite system reads
\be
\Xi_{\Lambda}= \Tr \exp[-\beta(H-\sum_{\alpha=1}^{\cS} \mu_{\alpha}N_{\alpha})] \; ,
\label{VI.48}
\ee
where the trace runs on all particle numbers, not only on neutral configurations. 
The grand canonical pressure
\begin{align}
P_\Lambda(T; \{ \mu_{\alpha} \})=\frac{k_{\rm B}T \ln \Xi_{\Lambda}}{\Lambda}
\label{VI.48bis}
\end{align}
has a  well-defined thermodynamic limit 
\begin{align}
P(T; \{ \mu_{\alpha} \})= k_{\rm B}T  \lim_{\rm \Lambda \to \infty}\frac{ \ln \Xi_{\Lambda}}{\Lambda} \; .
\label{PressureTL}
\end{align} 

\bigskip

As a consequence of elementary electrostatics, for non-neutral configurations
associated with $\sum_{\alpha=1}^{\cS} e_{\alpha}N_{\alpha} = Q \neq 0$, the excess charges are expelled to the surface~\cite{Lieb1972}, 
so the system maintains charge neutrality in the bulk. Moreover, the Coulomb energy associated with these excess charges 
is of order $Q^2/(2R)$ for a spherical box of radius $R$. Non-neutral configurations with a macroscopic charge proportional to the volume 
$\Lambda$ do not contribute to $\Xi$, since their weights involve the factor $\exp(-\beta Q^2/R)$ which 
vanishes faster than $\exp(- C \Lambda^{1+ \epsilon})$ with $C,\epsilon >0$. In fact,  $\langle Q \rangle_{\Lambda,\rm GC}$ remains 
of order $R$ in the TL  whatever the chemical potentials are, so the average charge density vanishes in the TL,  
\be
\la{AverageCharge}
\lim_{\rm TL} \frac{\langle Q \rangle_{\Lambda,{\rm GC}}}{ \Lambda} = 0,
\ee 
while the total surface charge-density carried by the walls of the box 
also vanishes in the TL~\cite{Lieb1972}. 
Hence the average densities defined by 
\be
\la{AverageDensity}
\rho_{\alpha}= \lim_{\rm TL} \frac{\langle N_\alpha \rangle_{\Lambda,\rm GC}}{\Lambda}
\ee
satisfy the overall charge neutrality
\begin{equation}
\sum_{\alpha=1}^{\cS}e_{\alpha}\rho_{\alpha}=0 \; .
\label{VI.49}
\end{equation}   
Moreover, they coincide with the local bulk densities in the TL in a fluid phase. 
Thanks to the thermodynamic identity
\be
\la{DensityChemicalPotential}
\rho_{\alpha}= \frac{\partial P}{\partial \mu_{\alpha}}(T;\{ \mu_{\gamma} \}) \; ,
\ee
charge neutrality~(\ref{VI.49}) can be recast as
\begin{equation}
\sum_{\alpha=1}^{\cS}e_{\alpha} \frac{\partial P}{\partial  \mu_{\alpha}} (T;\{ \mu_{\gamma} \})=0 \; .
\label{VI.50}
\end{equation}
The identity~(\ref{VI.50}) is valid for any set $\{ \mu_{\gamma} \}$. Because of this identity, 
the bulk properties do not depend independently on all $\cS$ chemical potentials: there is necessarily 
a combination of these chemical potentials that is irrelevant. This remarkable property allows one to 
introduce $(\cS -1)$ independent neutral-group chemical potentials, as well as the pseudo-neutrality condition~\eqref{PseudoNeutrality}, 
as detailed in the next sections~\ref{sec:II.B} and \ref{sec:II.Bbis}.

\subsection{Introduction of $(\cS - 1)$ independent Neutral-Group chemical potentials} 	
\label{sec:II.B}

We start with the simplest case of a two-component system ($\cS=2$) made with nuclei ($\alpha = 1={\rm n}$) carrying a charge $e_1= Z e$ and electrons ($\alpha=2={\rm e}$) carrying a charge $e_2=-e$.
Due to the identity~\eqref{VI.50}, there is one relevant combination of chemical potentials which entirely determines the equilibrium state in the TL.
As discussed previously the leading configurations which contribute to the grand-canonical trace~\eqref{VI.48} are almost neutral.
i.e. the numbers $(N_{\rm n}, N_{\rm e})$ of nuclei and electrons are such that $N_{\rm e} \simeq N_{\rm n} Z$. Accordingly, 
$(\mu_{\rm n} N_{\rm n} +\mu_{\rm e} N_{\rm n})$ is close to $\mu N_{\rm n}$ with
\be
\la{RelevantCombination}
\mu= \mu_{\rm n} + Z \mu_{\rm e} \; ,
\ee
which can be viewed as the chemical potential of an elementary neutral-group made with a single nuclei and $Z$ electrons. 
Such a linear combination, together with $T$, entirely determines the pressure,\textsl{ i.e.} 
$P(T; \mu_{\rm n},\mu_{\rm e})=P(T;\mu)$, { in agreement with the Lieb-Lebowitz theorem~\cite{Lieb1972,Brydges1999}}. The particle densities~(\ref{DensityChemicalPotential}) can then be recast as
\begin{align}
\la{DensityChemicalBis}
&\rho_{\rm n}= \frac{\partial P}{\partial \mu}(T;\mu)\frac{\partial \mu}{\partial \mu_{\rm n}}
=\frac{\partial P}{\partial \mu}(T;\mu) \nonumber \\
&\rho_{\rm e}= \frac{\partial P}{\partial \mu}(T;\mu)\frac{\partial \mu}{\partial \mu_{\rm e}} \; ,
=Z \frac{\partial P}{\partial \mu}(T;\mu) \; .
\end{align} 
and they obviously satisfy local charge neutrality.

\bigskip

For multi-component systems with three or more components, we can determine in a similar way $(\cS-1)$ relevant combinations of the chemical potentials.
Let us consider that species $(\alpha=1,...,\cS-1)$ are nuclei with charges $Z_\alpha e$, while species $\alpha=\cS={\rm e}$ are electrons with charges 
$-e$. Elementary neutral-groups can be constructed by associating $Z_\alpha$ electrons to a single given nuclei with species $\alpha$. 
The associated neutral-group (NG) chemical potentials are the $(\cS-1)$ combinations
\be
\la{RelevantCombinationGeneral}
\mu_\alpha^{\rm NG} = \mu_\alpha + Z_\alpha \mu_{\rm e} \, , \quad \alpha=1,...,\cS-1 \; , 
\ee 
which, together with the temperature, entirely determine the equilibrium state. Of course, when $\cS \geq 3$, there are several ways to constitute $(\cS-1)$ elementary neutral groups \footnote{For instance, one could define another set of neutral groups from $(\cS-1)$ vectors orthogonal to the charge vector $\pmb{e}=(Z_1, ..., Z_{\cS-1}, -1)e$ by setting the abundance of particles of species $\alpha$ in the neutral group associated to a vector $\vec{v}$ orthogonal to~$\pmb{e}$ to be proportional to the component $v_\alpha$ of that vector.
 The particular choice~\eqref{RelevantCombinationGeneral} for neutral-groups is associated with a particularly simple basis for the subspace orthogonal to $\pmb{e}$.}. This freedom of choice for the set of independent relevant variables $\{\mu_\alpha^{\rm NG}\}$ is however inconsequential in an exact calculation and it would not affect any physical prediction. This arbitrariness is due to the fact that there are several ways of grouping particles together such that each group is charge-neutral.

\subsection{Neutral-Group activities} 	\label{sec:II.Bbis}

It is useful to translate the previous considerations in terms of the particle activities
\be
z_\alpha=(2s_\alpha+1)\frac{\e^{\beta\mu_{\alpha}}}{(2\pi \lambda_\alpha^{2})^{3/2}}\; ,
\la{ParticleActivity}
\ee
where $\lambda_\alpha=(\beta \hbar^2/m_\alpha)^{1/2}$ is the de Broglie thermal wavelength of the 
particles of species $\alpha$. Let us consider the neutral-group chemical potentials~(\ref{RelevantCombinationGeneral}).
They provide $(\cS-1)$ neutral-group activities
\be
\la{RelevantActivity} 
y_i  =
\left[z_i z_{\rm e}^{{{Z}}_i} \right]^{1/(1+{Z}_i)}  
\ee
where the exponent $1/(1+Z_i)$ has been introduced in the definition of $y_i$ so that it has the dimension of an activity, \textit{i.e.} a density.  
The pressure depends solely on the $(\cS-1)$ neutral-group activities $y_i$ and on the temperature, i.e. 
$P(T;\{ \mu_{\alpha} \})=P(T;\{ \mu_{\alpha}^{\rm NG}\})=P(T;\{ y_i \})$. The thermodynamical identity~(\ref{DensityChemicalPotential}) 
which provides the particle densities is then rewritten as 
\be
\la{DensityNewActivity}
\rho_{\alpha}= z_\alpha \sum_{i=1}^{\cS-1} \frac{\partial \beta P}{\partial y_i}(T;\{ y_j \}) 
\frac{\partial y_i}{\partial z_{\alpha}}(\{ z_{\gamma} \})\; .
\ee
The total local charge density reads
\be
\la{ChargeDensityNewActivity}
\sum_\alpha e_\alpha \rho_{\alpha}=  \sum_{i=1}^{\cS-1} \frac{\partial \beta P}{\partial y_i}(T;\{ y_j \}) 
\sum_\alpha e_\alpha z_\alpha \frac{\partial y_i}{\partial z_{\alpha}}(\{ z_{\gamma} \})\; ,
\ee
and it indeed always vanishes since
\begin{equation}
\la{OrthogonalityProperty}
\sum_\alpha e_\alpha z_\alpha \frac{\partial y_i}{\partial z_{\alpha}}(\{ z_{\gamma} \})
= Z_i e z_i\frac{\partial y_i}{\partial z_i} -e z_{\rm e}\frac{\partial y_i}{\partial z_{\rm e}}  = e y_i \left(\frac{Z_i}{Z_i+1}-\frac{Z_i}{Z_i+1} \right) = 0\; .
\end{equation}

\bigskip

Since the particle densities are determined solely by the $(\cS-1)$
neutral-group activities $y_i$ and by the temperature, different sets of activities can lead to the same set of densities $\{\rho_\alpha\}$. It is common to break this redundancy by imposing, without any loss of generality
as far as bulk properties are concerned, the so-called pseudo-neutrality (or bare-neutrality) condition~\cite{Brydges1999,Brown2001}
\be
\la{PseudoNeutrality}
\sum_\alpha e_\alpha z_\alpha=0 \; .
\ee 
{Notice that fixing the electrons' activity in terms of the $(\cS-1)$ nuclei's activities via this relation does not affect the range of variations $[0, \infty]$ of each $y_i$ variable.}
The choice~(\ref{PseudoNeutrality}) is particularly useful for various purposes, in particular it simplifies the derivation of the 
low-density expansion of the EOS as explained in Section~\ref{sec:S3}.

\subsection{Neutral-Group neutralization scheme}
\label{sec:S23}

\subsubsection{NG neutralization prescription}

A given approximate theory, that is a given function  for the pressure $P_{\rm A}(T; \{z_\alpha\})$ in the TL, is not necessarily 
compatible with neutrality, {\it i.e.} the particle densities inferred \textsl{via } the standard identities
\be
\la{DensityPressureA}
\rho_{\alpha}= z_{\alpha} \frac{\partial P_{\rm A}}{\partial z_{\alpha}}(T;\{ z_{\gamma} \}) \; ,
\ee
do not satisfy the local charge neutrality~(\ref{VI.49}) in general. In other words, 
it is not possible to express $P_{\rm A}(T; \{z_\alpha\})$ solely in terms of the neutral-group activities $\{y_i\}$ and the temperature, 
as it can be done for the exact pressure. However, one can modify the approximate theory via the following general procedure 
to make it compatible with neutrality. 

\bigskip

Let us introduce the associated approximation 
\be
\la{NGpressure}
P_{\rm A}^{\rm NG}(\beta;\{ z_\alpha \})=P_{\rm A}(\beta;\{ z_\alpha^{\rm NG}(y_1(\{z_\alpha\}),...,y_{\cS-1}(\{z_\alpha\})) \})
\ee
where each $z_\alpha$ in $P_{\rm A}(\beta;\{ z_\alpha\})$ is replaced by a Neutral-Group function 
$z_\alpha^{\rm NG}(y_1(\{z_\alpha\}),$ $..., y_{\cS-1}(\{z_\alpha\}))$ which depends on the genuine activities $\{z_\alpha\}$ 
through the neutral-group activities $\{y_i\}$ [Eq.~\eqref{RelevantActivity}].
The dependence of the NG functions $\{z_\alpha^{\rm NG}\}_{\alpha=1, ..., \cS}$ on the variables $\{y_i\}_{i, ..., \cS-1}$ 
is obtained by inverting the system of equations
\begin{subequations}
	\begin{numcases}{}
y_i = \left[z_i^{\rm NG} (z_{\rm e}^{\rm NG})^{{{Z}}_i} \right]^{1/(1+{Z}_i)}, \qquad i = 1, ..., \cS-1	\label{29a} \\
\sum_{\alpha=1}^\cS e_\alpha  z_\alpha^{\rm NG} = 0 \hspace{5.2cm}\mbox{}
		\label{29b}
	\end{numcases}
\end{subequations}
which combines the definitions~\eqref{RelevantActivity} of the neutral-group variables [where $z_i$ is replaced by $z_i^{\rm NG}$] with pseudo-neutrality. 
Hence, for the specific set of genuine activities $\{z_\alpha\}$ which satisfy pseudo-neutrality, 
each function $\{ z_\alpha^{\rm NG}(\{y_i\}) \}$ takes the value $\{z_\alpha\}$. Notice that the variations of the functions 
$\{z_\alpha^{\rm NG}\}$, with the $\{z_\alpha\}$'s treated as independent variables, are entirely defined by the choice of 
neutral groups. The Neutral-Group functions do not depend in particular on the considered approximate theory.
The present ``back-and-forth'' conversion, from the $\cS$ genuine activities to $(\cS-1)$ neutral-group activities $\{y_i\}$ 
to $\cS$ activity-functions $\{z_\alpha^{\rm NG}(z_{\gamma})\}$, ensures that the associated approximation $P_{\rm A}^{\rm NG}$ 
depends on the activities only via the neutral-group activities, and therefore that $\Omega^{\rm NG} = -P_{\rm A}^{\rm NG}(T,\{z_\alpha\})\Lambda$ is a thermodynamic potential compatible with electroneutrality.

\bigskip

By construction, the associated pressure $P_{\rm A}^{\rm NG}(\beta;\{ z_\alpha \})$ only depends on the activities $\{z_\alpha\}$ via the relevant neutral-group activities, so it leads to particle densities that satisfy local charge neutrality.
The particle densities can be computed by applying 
the general rules for partial derivatives of composite functions, which provides
\begin{align}
\la{DensityPressureANG}
\rho_{\alpha} &= z_{\alpha} \frac{\partial P_{\rm A}^{\rm NG}}{\partial z_{\alpha}}(T;\{ z_{\gamma} \}) 
= z_{\alpha} \frac{\partial }{\partial z_{\alpha}} P_{\rm A}(T;\{ z_\gamma^{\rm NG}(y_1(\{z_\delta\}),...,y_{\cS-1}(\{z_\delta\})) \})
 \nonumber \\
&= z_{\alpha} \sum_{\delta=1}^{\cS} \sum_{i=1}^{\cS-1} \frac{\partial P_{\rm A}}{\partial z_{\delta}} (T;\{z_\gamma^{\rm NG} \}) 
\frac{\partial z_{\delta}^{\rm NG}}{\partial y_i} 
\frac{\partial y_i}{\partial z_{\alpha}}\; ,
\end{align}
The partial derivatives $\partial P_{\rm A}/\partial z_{\gamma}$, with $z_\gamma$ treated as an independent variable, have to be evaluated 
at the end for the set $\{ z_\gamma^{\rm NG}(y_1,...,y_{\cS-1}) \}$
that satisfies the pseudo-neutrality condition~(\ref{PseudoNeutrality}). 
It is convenient to consider that the set of genuine activities $\{ z_\gamma \}$ already 
satisfy this condition since each function $z_\gamma^{\rm NG}(y_1,...,y_{\cS-1})$ then exactly coincides with $z_\gamma$ at the end. 

\bigskip

With this neutralization prescription, the pressure is left unchanged for a pseudo-neutral set $\{z_\alpha\}$, 
\textit{i.e.} $P_{\rm A}^{\rm NG} = P_{\rm A}$ for such set, and the theory is internally consistent since the densities are deduced  from the pressure via the standard thermodynamic relation $\rho_\alpha = z_\alpha \frac{\partial}{\partial z_\alpha} P_{\rm A}^{\rm NG}(T,\{ z_\gamma\})$.

\subsubsection{Explicit neutralization formulae}

Using the definition~(\ref{RelevantActivity}) of the Neutral-Group activities, the densities~(\ref{DensityPressureANG}) 
can be recast as 
\begin{align}
\la{DensityPressureANGbis}
&\rho_i =  \sum_{\theta=1}^{\cS} \frac{\partial P_{\rm A}}{\partial z_{\theta}} (T;\{z_\gamma \}) 
\frac{y_i}{Z_i+1} \frac{\partial z_{\theta}^{\rm NG}}{\partial y_i} 
\quad \text{for} \quad i=1,...,\cS-1 \nonumber \\
&\rho_{\rm e}= \sum_{\theta=1}^{\cS} \frac{\partial P_{\rm A}}{\partial z_{\theta}} (T;\{z_\gamma \})  
\sum_{j=1}^{\cS-1} \frac{Z_j y_j}{Z_j+1} \frac{\partial z_{\theta}^{\rm NG}}{\partial y_j}
\; .
\end{align}
Taking the logarithm of the definitions~(\ref{RelevantActivity}), we obtain  
\be
\la{RelevantActivityLogarithm} 
(Z_i+1) \ln y_i  = \ln z_i^{\rm NG} + Z_i \ln z_{\rm e}^{\rm NG}  \quad i=1,...,\cS-1 \; .
\ee
The partial derivatives ${\partial z_{\theta}^{\rm NG}}/{\partial y_i}$ can be calculated by differentiating each side of Eq.~(\ref{RelevantActivityLogarithm}) 
with respect to $y_i$ in a first step, and then with respect to $y_j$ for $j \neq i$ in a second step. Inserting the resulting 
expressions for ${\partial z_{\theta}^{\rm NG}}/{\partial y_i}$ into the 
formula~(\ref{DensityPressureANGbis}), we eventually find
\begin{align}
\la{DensityPressureANGter}
&\rho_i =  \sum_{\theta=1}^{\cS} C_{i,\theta} \; z_\theta \; \frac{\partial P_{\rm A}}{\partial z_{\theta}} (T;\{z_\gamma \}) 
 \quad \text{for} \quad i=1,...,\cS-1 \nonumber \\
&\rho_{\rm e}=   \sum_{j=1}^{\cS-1} Z_j \rho_j \; ,
\end{align} 
with coefficients 
\begin{equation}
\la{CoefficientsNGDensityA} 
C_{i,\theta}=
\begin{cases}
\displaystyle
- \frac{Z_i  Z_j z_i}{2 z_{\rm e} + Z_i(Z_i-1)z_i}    & \text{if $\theta \neq i$ and $\theta \neq \cS$} \\
\displaystyle\rule{0mm}{8mm}
1-\frac{Z_i^2 z_i}{2 z_{\rm e} + Z_i(Z_i-1)z_i}  & \text{if $\theta = i$} \\
\displaystyle\rule{0mm}{8mm}
 \frac{Z_i  z_i}{2 z_{\rm e} + Z_i(Z_i-1)z_i}   & \text{if $\theta = S$ [\textsl{i.e.} $\theta$ = e]}
\end{cases}
\end{equation}

\bigskip

By construction, the particle densities generated by the Neutral-Group prescription \eqref{DensityPressureANGter}-\eqref{CoefficientsNGDensityA} 
do satisfy the local charge 
neutrality~(\ref{VI.49}), whatever the partial derivatives $\partial P_{\rm A}/\partial z_{\gamma}$ are. We recall that 
such derivatives must be calculated for a set $\{ z_\gamma \}$ which fulfills the pseudo-neutrality condition~(\ref{PseudoNeutrality}).

\bigskip

As a first check, we verify that if the approximate pressure $P_{\rm A}$ is consistent with local charge neutrality, namely if
\be
\la{APconsistencyNeutrality}
 z_{\rm e} \frac{\partial P_{\rm A}}{\partial z_{\rm e}}  = \sum_{j=1}^{\cS-1} Z_j z_j \frac{\partial P_{\rm A}}{\partial z_j} \; , 
\ee 
then formula~(\ref{DensityPressureANGter}) for each density $\rho_i$ does reduce to $z_i\partial P_{\rm A}/\partial z_{i} $ by virtue of the 
identities
\begin{align}
\la{CoefficientsIdentity} 
& Z_j C_{i,{\rm e}} + C_{i,j}= 0  \quad \text{for} \quad j=1,...,\cS-1 \; , \; j \neq i \nonumber \\
& Z_i C_{i,{\rm e}} + C_{i,i}= 1 \; .
\end{align} 
Moreover, if the approximate pressure is the ideal Maxwell-Boltzmann expression, i.e. if
\be
\la{MBPressure}
\beta P_{\rm A}= \beta P_{\rm MB}= z_{\rm e}+\sum_{j=1}^{\cS-1} z_j
\ee 
each density $\rho_i$ given by formula~(\ref{DensityPressureANGter}) does reduce to $z_i$ as a consequence of  
$
\sum_{j=1}^{\cS-1}  C_{i,j} z_j + C_{i,{\rm e}} z_{\rm e}= z_i 
$.

\subsubsection{Explicit formulae for two- and three-component plasmas}

Let us consider a two-component plasma, $\cS=2$ with $(1={\rm n},Z_1=Z$).
The nuclei and electron densities~(\ref{DensityPressureANGter}) then become
\begin{align}
\la{DensityPressureANGTCP}
&\rho_{\rm n} =  \frac{z}{(Z+1)}
\left[ \frac{\partial P_{\rm A}}{\partial z_{\rm n}}   + Z \frac{\partial P_{\rm A}}{\partial z_{\rm e}} \right]
 \nonumber \\
&\rho_{\rm e}= \frac{Z z}{(Z+1)}
\left[ \frac{\partial P_{\rm A}}{\partial z_{\rm n}}   + Z \frac{\partial P_{\rm A}}{\partial z_{\rm e}} \right]
\; ,
\end{align}
where $\partial P_{\rm A}/\partial z_{\rm n}$ and $\partial P_{\rm A}/\partial z_{\rm e}$ are calculated for 
the set $(z_{\rm n}=z,z_{\rm e}=Z z )$. Note that for the hydrogen plasma, the nuclei are protons with $Z=1$, while for 
the helium plasma the nuclei are alpha-particles with $Z=2$.

\bigskip

For three-component systems, $\cS=3$, the nuclei densities~(\ref{DensityPressureANGter}) read
\begin{multline}
\la{DensityPressureANG3CP1}
\rho_1 =  \frac{(Z_1z_1 +2Z_2 z_2) z_1 }{Z_1(Z_1+1) z_1 + 2Z_2 z_2}  \; \frac{\partial P_{\rm A}}{\partial z_1} (T;\{z_\gamma \}) 
-\frac{Z_1 Z_2 z_1 z_2}{Z_1(Z_1+1) z_1 + 2Z_2 z_2}  \frac{\partial P_{\rm A}}{\partial z_2} (T;\{z_\gamma \}) \\
+ \frac{Z_1  z_1 (Z_1z_1+Z_2z_2)}{Z_1(Z_1+1) z_1 + 2Z_2 z_2} \frac{\partial P_{\rm A}}{\partial z_{\rm e}} (T;\{z_\gamma \}) \; ,
\end{multline} 
and 
\begin{multline}
\la{DensityPressureANG3CP2}
\rho_2 =   \frac{(2Z_1z_1 + Z_2 z_2)z_2}{2 Z_1 z_1 + Z_2(Z_2+1)z_2}  \; \frac{\partial P_{\rm A}}{\partial z_2} (T;\{z_\gamma \}) 
-\frac{Z_1 Z_2 z_1 z_2}{2 Z_1 z_1 + Z_2(Z_2+1)z_2} \;  \frac{\partial P_{\rm A}}{\partial z_1} (T;\{z_\gamma \}) \\
+ \frac{Z_2  z_2(Z_1z_1+Z_2z_2)}{2 Z_1 z_1 + Z_2(Z_2+1)z_2} \; \frac{\partial P_{\rm A}}{\partial z_{\rm e}} (T;\{z_\gamma \})\; ,
\end{multline}
with the electron density $\rho_{\rm e}=Z_1 \rho_1 + Z_2 \rho_2$. These formulae can be applied to the case of 
the hydrogen-helium mixture made with protons $(1={\rm p},Z_1=1$) and alpha-nuclei ($2={\rm alpha}, Z_2=2$). The proton and alpha-particle activities 
$z_{\rm p}$ and $z_{\rm alpha}$ can take arbitrary values, while the electron activity is set to $z_{\rm e}=z_{\rm p} + 2 z_{\rm alpha}$. 
The nuclei-densities depend on the two independent 
activities $z_{\rm p}$ and $z_{\rm alpha}$. Note that the relative concentrations of hydrogen and helium, determined 
by the ratio $\rho_{\rm p}/\rho_{\rm alpha}$, do not depend only on the ratio $z_{\rm p}/z_{\rm alpha}$ in general.

\subsubsection{Comments}

The present neutral-group neutralization prescription if quite appealing because: 
(i) It is general and straightforward to implement since no equation needs to be solved; 
(ii) It is based on exact properties of the system, namely that it maintains neutrality in the bulk and that the dependence 
of the pressure on the activities occurs only via $(\cS-1)$ neutral-group activities $\{y_i\}$ which have a clear physical interpretation; 
(iii) The associated electroneutrality-compatible theory $P_{\rm A}^{\rm NG}(T,\{z_\alpha\})$ 
is internally consistent since the densities are deduced from this function via the standard thermodynamic relation; 
(iv) The original value of the pressure is left unchanged after neutralization $P_{\rm A}^{NG} = P_{\rm A}$ if the genuine $z_\alpha$'s satisfy the pseudo-neutrality condition.

\bigskip

The neutral-group neutralization prescription is not unique in a plasma with 3 or more components. Indeed, when $\cS \geq 3$, 
other choices of the neutral groups would lead to expressions of the particle densities 
similar to Eqns.~(\ref{DensityPressureANGter}) but with coefficients different from Eqns.~(\ref{CoefficientsNGDensityA}).
However, the formulae for these coefficients in terms of the particle activities are expected to be much more complicated than the 
rational fractions~(\ref{CoefficientsNGDensityA}). In fact the choice~(\ref{RelevantActivity}) ensures that 
$\partial y_i /\partial z_j = 0$ for $j \neq i$, which greatly simplifies the calculations of the $C_{i,\delta}$'s.

\section{Activity expansion of the pressure}	\label{sec:S3}

Mayer diagrams have been introduced while ago~\cite{Mayer1940} in order to derive 
low-density expansions of equilibrium quantities for classical 
systems with short-range pair interactions. For charged fluids, every Mayer diagram 
diverges because of the long-range of Coulomb interactions. 
Abe~\cite{Abe1959} and Meeron~\cite{Meeron1958} showed that such divergences can be removed \textit{via}
systematic summations of convolution chains built with the Coulomb interaction. The whole Mayer series is then 
exactly transformed into a series of so-called prototype graphs, with the same topological structure as the Mayer diagrams, but 
with effective bonds built with the familiar Debye potential in place of the bare Coulomb interaction. The contribution of each 
prototype graph is finite thanks to the screening  collective effects embedded in the Debye potential.

\subsection{The equivalent classical gas of loops}

The trace~(\ref{VI.48}) defining 
$\Xi_{\Lambda}$ can be expressed in position and spin space, where the corresponding states have to be symmetrized 
according to Bose or Fermi statistics. The corresponding sum involves both diagonal and off-diagonal 
matrix elements of $\exp(-\beta H_N)$ in position space. Diagonal matrix elements account for Maxwell-Boltzmann 
statistics, while off-diagonal matrix elements describe exchange contributions.
Within the Feynman-Kac representation, 
all the matrix elements of $\exp(-\beta H_N)$ in position space can be rewritten as 
functional integrals over paths followed by the particles. 
The off-diagonal matrix elements generate open paths. However all the open paths followed by 
the particles exchanged in a given cyclic permutation, can be collected into a closed filamentous object, 
called a loop $\cL$, or sometimes a ring-polymer, in the literature. 
Each contribution of a given spatial matrix element of $\exp(-\beta H_N)$ for a given set of particles can be 
related to that of a classical Boltzmann factor for a set of loops. In a last non-trivial step, the sum 
of all these contributions, namely $\Xi_{\Lambda}$, is recast as the grand-partition function of 
a classical gas of loops~\cite{Ginibre1971,Cornu1996,Martin2003}
\be
\Xi_{\Lambda}=\Xi_{\Lambda,{\rm Loop}}=\sum_{N=0}^{\infty}\frac{1}{N\;!}\[ \prod_{i=1}^{N} \int_{\Lambda}D(\cL_{i})z(\cL_{i})\]
\e^{-\beta U(\cL_{1},\cL_{2},\ldots,\cL_{N})} \;,
\la{IX.QMG6}
\ee
thereby establishing a mapping, at equilibrium, between the quantum gas and a classical gas of loops.
The loop phase-space measure $D(\cL)$, loop fugacity $z(\cL)$, and total interaction potential 
$U(\cL_{1},\cL_{2},\ldots,\cL_{N})$ are defined as follows.

\bigskip
 
 \begin{figure}[h]
 \includegraphics[scale=0.8]{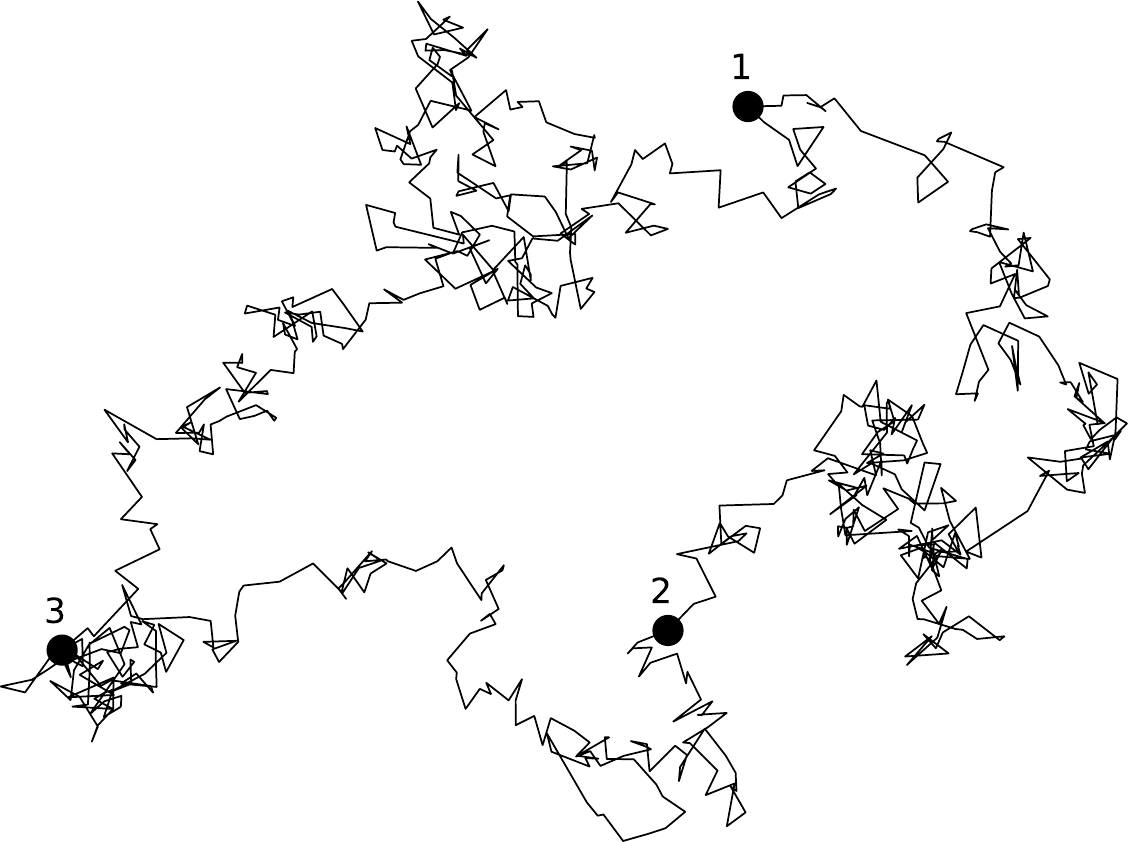}
 \caption{A loop made with the paths of 3 particles exchanged in a permutation cycle ($1 \to 2 \to 3 \to 1$).}
 \label{Fig1}
 \end{figure}
 
A loop $\cL$ located at $\bx$ containing $q$ particles of species $\alpha$, 
is a closed path $\bX(s)=\bx+\lambda_{\alpha}\bcX(s)$, parametrized by  an 
imaginary time $s$ running from $0$ to $q$ where $\bcX(s)$, the shape of the loop, is a Brownian bridge 
subjected to the constraints $\bcX(0)=\bcX(q)=\mathbf{0}$ {{(Fig.~\ref{Fig1})}}. 
The state of a loop, collectively denoted by $\cL=\{\bx, \chi\}$, is defined by its position $\bx$ together with an 
internal degree of freedom $\chi=\{\alpha, q,\bcX\}$, which includes its shape $\bcX$ as well as 
the number $q$ of exchanged particles of species $\alpha$. The loop phase-space measure $D(\cL)$ means summation over all these 
degrees of freedom,
\begin{equation}
\int_{\Lambda} D(\cL)\cdots=\sum_{\alpha=1}^{\cS}\sum_{q=1}^{\infty}\int_{\Lambda}\d \bx\int D_{q}(\bcX)\cdots \; .
\label{L.27}
\end{equation}
The functional integration over the loop shape $D_{q}(\bcX)$  is the normalized Gaussian measure for 
the Brownian bridge $\bcX(s)$ entirely defined by its covariance
\be
\int D_{q}(\bcX)\bcX_{\mu}(s_{1})\bcX_{\nu}(s_{2})=q\delta_{\mu\nu}\[\min\left(\frac{s_{1}}{q},\frac{s_{2}}{q}\right)-\frac{s_{1}}{q}\frac{s_{2}}{q}\] \; .
\la{IX.QMG7bis}
\ee
%
%
The loop activity reads
\be
z(\cL)=(2s_\alpha+1)\frac{\eta_\alpha^{q-1}}{q}\frac{\e^{\beta\mu_{\alpha}q}}{(2\pi q\lambda_\alpha^{2})^{3/2}}\e^{-\beta U_{\rm self}(\cL)} \; ,
\la{IX.QMG8}
\ee
where the factor $\eta_\alpha=1$ for bosons and $\eta_\alpha=-1$ for fermions.
Moreover, $U_{\rm self}(\cL)$ 
is the self-energy of the loop which is generated by the interactions between the exchanged particles,  
\begin{align}
U_{\rm self}(\cL)=\frac{e_\alpha^2}{2}\int_{0}^{q}\d s\int_{0}^{q}\d s'(1-\delta_{[s][s']})\tilde{\delta}(s-s')
v_{\rm C}(\lambda_\alpha \bcX(s)-\lambda_\alpha \bcX(s')) \;,
\label{IX.QMG9}
\end{align}
with the Dirac comb 
\begin{equation}
\tilde{\delta}(s-s')=\sum_{n=-\infty}^{\infty}\delta(s-s'-n)=\sum_{n=-\infty}^{{\infty}}\e^{2i\pi n(s-s')} \; .
\label{IX.QMG10}
\end{equation}
The Dirac comb ensures that particles only interact at equal times $s$ along their paths, as required by the Feynman-Kac formula, 
while the term $(1-\delta_{[s][s']})$ removes the contributions of self-interactions ($[s]$ denote the integer part of~$s$). 


Eventually, the total interaction potential $U(\cL_{1},\cL_{2},\ldots,\cL_{N})$ is a sum of pairwise interactions,
\be
\la{IX.QMG11}
U(\cL_{1},\cL_{2},\ldots,\cL_{N})=\frac{1}{2} \sum_{i \neq j} V(\cL_i,\cL_j)
\ee 
with 
\begin{align}
&V(\cL_i,\cL_j)=e_{\alpha_i} e_{\alpha_j} \int_{0}^{q_i}\d s_i\int_{0}^{q_j}\d s_j \tilde{\delta}(s_i-s_j)
v_{\rm C}(\bx_i+\lambda_{\alpha_i}\bcX_i(s_i)-\bx_j-\lambda_{\alpha_j}\bcX_j(s_j)) \; .
\label{IX.QMG12}
\end{align}
The loop-loop interaction $V(\cL_i,\cL_j)$ is generated by the interactions between any particle inside $\cL_i$ and any particle inside $\cL_j$. 
Like in formula~(\ref{IX.QMG9}), the Dirac comb~(\ref{IX.QMG10}) guarantees that interactions are taken at equal times along particle paths.

\bigskip

The introduction of the gas of loops is particularly useful at low densities, because the standard Mayer diagrammatic expansions, valid 
for classical systems with pairwise interactions, 
can be straightforwardly applied by merely replacing points by loops. However, as in the case of classical Coulomb systems, the Mayer diagrams for the loop gas are plagued with divergences arising from the large-distance behavior
\be
\la{IX.QMG28}
V(\cL_{i},\cL_{j}) \sim\frac{q_{i} e_{\alpha_{i}} q_{j} e_{\alpha_{j}}}{|\bx_{i}-\bx_{j}|} \;\quad \text{when} \; |\bx_{i}-\bx_{j}| \to \infty \; .
\ee
Note that such behavior is nothing but the Coulomb interaction between point charges, because the finite spatial extents of loops $\cL_i$ 
and $\cL_j$ can be neglected with respect to their large relative distance $|\bx_{i}-\bx_{j}|$. It has been shown that all these long-range 
divergences can be removed within a suitable extension of the Abe-Meeron summation process 
introduced long ago for classical Coulomb fluids. The method has been applied for both the one- and two-body distribution functions~\cite{Cornu1996,Alastuey2003}.
In the next Section, we derive the corresponding Abe-Meeron series for the pressure.

\subsection{Abe-Meeron like summations for the pressure}

The Mayer diagrammatical expansion of the pressure,
\begin{equation}
\la{VBMayerP}
\beta P =
\sum_{{\cal G}} \frac{1}{S({\cal G})} 
\int \left[ \prod D(\cL) z(\cL) \right] 
\left[ \prod b_{\rm M}\right]_{{\cal G}} \;,
\end{equation}
 involves simply connected diagrams ${\cal G}$ made with $N=1, 2, ...$ field (black) points, representing loops with statistical weight $z(\cL)$, and Mayer bonds $b_{\rm M}$ defined by
 \be
\la{MayerBond}
b_{\rm M}(\cL_i,\cL_j)= \exp(-\beta V(\cL_i,\cL_j)) -1 \; .
\ee
The contribution of a given ${\cal G}$ is 
calculated by labeling arbitrarily the $N$ field points (loops). $S({\cal G})$ denotes the symmetry factor, which is  the number of permutations of those labeled field loops that leave the product of bonds 
and weights unchanged.
An integration \eqref{L.27} is performed over the degrees of freedom of each field loop. 
Thanks to translation invariance, once the 
integration over $(N-1)$ black loops have been performed in ${\cal P}$, the result no longer depends on the position of the remaining black loop. The 
$1/\Lambda$ factor in the definition~(\ref{VI.48bis}) of the pressure of the finite system can then be absorbed in the thermodynamic limit
by keeping the position of one loop fixed, i.e. by integrating only over $(N-1)$ loops and on the internal degrees of freedom of the fixed loop.

\bigskip

Due to the large-distance behavior~\eqref{IX.QMG28}, any Mayer diagram ${\cal G}$ involving more than one loop is divergent in the thermodynamic limit. Let us eliminate these divergences systematically by summing diagrams in classes, as in the classical case~\cite{Abe1959,Meeron1958}. Since exactly the same counting and combinatorics formulae intervene in these summations as in the classical case, we won't detail them. Note that 
simplified presentations of the summation process for the one-body loop density are given in Refs.~\cite{Cornu1996} and~\cite{Alastuey2003}. 
The key staring point is the decomposition of the Mayer bond \eqref{MayerBond}
into
\be
\la{DecompositionBond}
b_{\rm M}(\cL_i,\cL_j)=b_{\rm T}(\cL_i,\cL_j) + b_{\rm I}(\cL_i,\cL_j) \; ,
\ee
with the interaction bond
\be
\la{InteractionBond}
b_{\rm I}(\cL_i,\cL_j)=-\beta  V(\cL_i,\cL_j) \; 
\ee
and the truncated bond
\be
\la{TruncatedBond}
b_{\rm T}(\cL_i,\cL_j)= \exp(-\beta V(\cL_i,\cL_j)) -1 + \beta  V(\cL_i,\cL_j) \; .
\ee
\begin{figure}[h]
\includegraphics[scale=1.15]{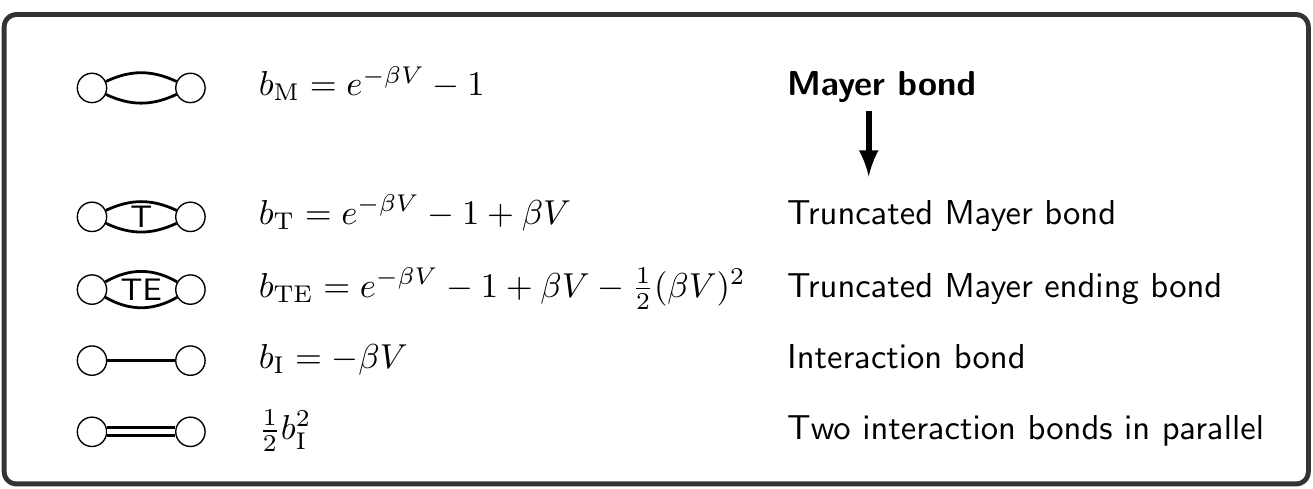}
\caption{The bonds before summations. The last four bonds are generated by decomposing the original Mayer bond.}
\label{FigBareBonds}
\label{Fig2}
\end{figure}
\begin{figure}[h]
\includegraphics[scale=1.15]{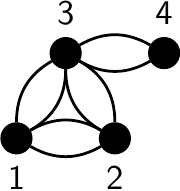}
\caption{Diagram with an ending loop (point 4).}
\label{Fig3}
\end{figure}
Graphical representations for these bonds are given in Fig.~\ref{FigBareBonds}.
A loop $\cL_i$ which is singly connected to a loop $\cL_j$ is called an ending loop (Fig.~\ref{Fig3}).
A Mayer bond $b_{\rm M}(\cL_i,\cL_j)$ connected to such a loop is decomposed as 
\be
\la{DecompositionBondEnding}
b_{\rm M}(\cL_i,\cL_j)=b_{\rm TE}(\cL_i,\cL_j) + b_{\rm I}(\cL_i,\cL_j) + [b_{\rm I}(\cL_i,\cL_j)]^2/2\; ,
\ee
with the truncated ending bond
\be
\la{TruncatedEndingBond}
b_{\rm TE}(\cL_i,\cL_j)=\exp (-\beta V(\cL_i,\cL_j)) - 1 + \beta V(\cL_i,\cL_j)-(\beta V(\cL_i,\cL_j))^2/2 \; .
\ee
These two decompositions can be represented graphically
\begin{align}
&\includegraphics{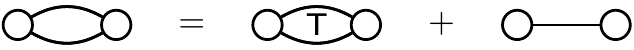}\\
&\includegraphics{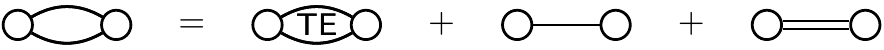}\;.
\end{align}
After inserting these decompositions into every diagram ${\cal G}$, a pair of loops $\cL_k$ and $\cL_l$ can be connected either by $b_{\rm I}$ or $b_{\rm T}$ (if none of the two loops is an ending loop) or by $b_{\rm I}$, $\frac{1}{2} b^2_{\rm I}$ or $b_{\rm TE}$ (if at least one of the two loops is an ending loop). We proceed then to systematic summations of all chain convolutions $b_{\rm I} \ast b_{\rm I} \ast...b_{\rm I} \ast b_{\rm I} $ made with arbitrary numbers $p$
of interaction bonds $b_{\rm I}$. Such a convolution chain can link a loop $\cL_i$ to another loop $\cL_j$ or to itself ($\cL_j = \cL_i$), in which case we call this convolution chain a ring.
%
%
The sum of $p=1,2,... \infty$ single convolution chains between two fixed loops $\cL_i$ and~$\cL_j$,
\be	\label{defPhi}
\begin{graph}
\node[label=below:{$i$},blanc] (1) {};
\node[label=below:{$j$},blanc] [right of=1] (2) {};
\drawLinkD{1}{2} ;
\end{graph}
\quad = \quad
\begin{graph}
\node[blanc] (1) {};
\node[blanc] [right of=1] (2) {};
\drawLinkFC{1}{2} ;
\end{graph}
\;\, + \;\,
\begin{graph}
\node[blanc] (1) {};
\node[noir] [right of=1] (2) {};
\node[blanc] [right of=2] (3) {};
\drawLinkFC{1}{2};
 \drawLinkFC{2}{3} ;
\end{graph}
\;\, + \;\,
\begin{graph}
\node[blanc] (1) {};
\node[noir] [right of=1] (2) {};
\node[noir] [right of=2] (3) {};
\node[blanc] [right of=3] (4) {};
 \drawLinkFC{1}{2} ;
 \drawLinkFC{2}{3} ;
 \drawLinkFC{3}{4} ;
\end{graph}
\;\, + \;\,
 \cdots
\ee
generates the Debye bond
\be
\la{DebyeBond} 
b_{\rm D}(\cL_i,\cL_j)=-\beta e_{\alpha_i} e_{\alpha_j} \phi(\cL_i,\cL_j) \; ,
\ee
where $\phi(\cL_i,\cL_j)$  is the quantum analogue of the Debye potential, which reads~\cite{Ballenegger2002} 
\be
\la{IX.QMG47}
\phi(\cL_{i},\cL_{j})= \int_{0}^{q_i}\d s_i\int_{0}^{q_j}\d s_j  
\; \psi_{\rm loop}(\bx_j+\lambda_{\alpha_{j}}\bcX_j(s_j)-\bx_i-\lambda_{\alpha_i}\bcX_i(s_i), s_i-s_j) \; ,
\ee
with
\be
\la{IX.QMG48}
\psi_{\rm loop}(\br,s)=  \sum_{n=-\infty}^\infty \exp(2 i \pi n s) \tilde{\psi}_{\rm loop}(\br,n) \; \; 
\ee
and
\be
\la{IX.QMG49}
\tilde{\psi}_{\rm loop}(\br,n)= \int \frac{ \d^3 \bk}{(2\pi)^3} \exp(i \bk \cdot \br)   \frac{4 \pi }{k^2 + \kappa^2(k,n)} \; .
\ee
Note that $\tilde{\psi}_{\rm loop}(\br,n)$ has a structure analogous to the classical Debye form, 
except that an infinite number of frequency-dependent screening factors $\kappa^{2}(k,n)$ occur, 
\be
\la{IX.QMG45a}
\kappa^2(k,n)=4\pi \beta \sum_\alpha \sum_{q=1}^\infty q e_{\alpha}^2 \int_0^q \d s \exp(2 i \pi n s) \int D_{q}(\bcX)
\exp(i \bk \cdot \lambda_{\alpha}\bcX(s)) z(\chi) \; .
\ee
The collective effects are embedded in these screening factors $\kappa^2(k,n)$, while the frequencies $2 \pi n$ are the analogues 
of the familiar Matsubara frequencies in the standard many-body perturbative series.

\bigskip

Similarly to the case of the Mayer diagrams for the one-body loop density, the summation of all convolution chains in the Mayer 
diagrams for the pressure can be expressed in terms of $\phi$, except in the single ring diagrams  built with arbitrary numbers $p \geq 2$ of interaction bonds $b_{\rm I}$, 
\begin{equation}	\label{eq:rings}
	\beta P_{\rm R} = \quad
	\begin{graph}
		\twoNodesUnNamed;
		\drawLinkFCC{1}{2};
	\end{graph}
	\,+\,
	\begin{graph}
		\ringsAttached{3}{0.6}{1};
	\end{graph}
	\,+\,
	\begin{graph}
		\ringsAttached{4}{0.6}{1};
	\end{graph}
	\,+\,
	\begin{graph}
		\ringsAttached{5}{0.6}{1};
	\end{graph}
	\,+\,\ldots 
\quad\equiv \begin{graph}
	\node[noir, thick] (1) {};
	\drawLoopNsnakeAngle{1}{90};
	\end{graph}
\end{equation}
which provide the contribution $\beta P_{\rm R}$.
In such diagrams made with $p$ black points, the symmetry factor is $1/(2p)$, in contrast to the single chain diagrams in Eq.~\eqref{defPhi} where the symmetry factor is $1$ for any $p$. 
After expressing each bare interaction $V$ in Fourier space, we find that the contribution to the pressure of a 
single ring made with $p$ black loops and $p$ bonds $b_{\rm I}=-\beta V$ reduces to
\be
\la{RingPp}
\frac{1}{2}\sum_{n=-\infty}^\infty \int \frac{\d^3 \bk}{(2\pi)^3}\frac{1}{p} \left[\frac{-\kappa^2(k ,n)}{k^2}\right]^p \; .
\ee
The calculation is similar to that involved in the convolution chain and gives again rise to the screening factors $\kappa^{2}(k,n)$.
Now, the summation over $p$ of all ring contributions leads to a logarithmic function instead of the rational fraction $1/[k^2 + \kappa^{2}(k,n) ]$ for
the chain contributions, namely
\begin{equation}
\label{RingSum}
\beta P_{\rm R} = \frac{1}{2}\sum_{n=-\infty}^\infty  \int \frac{\d^3 \bk}{(2\pi)^3} \left[
\frac{\kappa^2(k ,n)}{k^2} - \ln\left(1+ \frac{\kappa^2( k,n)}{k^2}\right) \right] \; .
\end{equation}

\bigskip

\begin{figure}
\vskip 8mm
\includegraphics[scale=1.15]{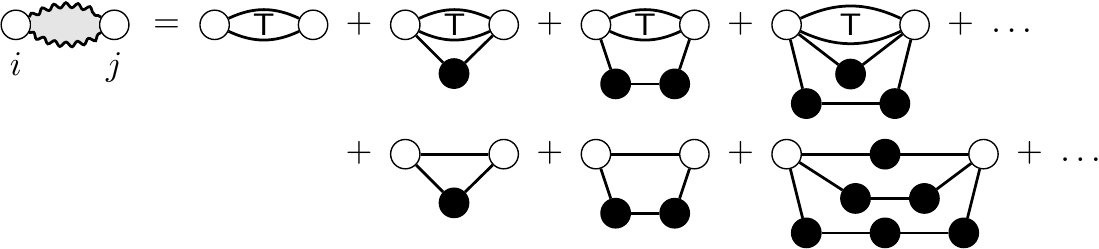}
\caption{Examples of diagrams contributing to the bond $b_{\rm AM}(\cL_i,\cL_j)$. Since the truncated Mayer bond $b_{\rm T}$ can be interpreted as the sum of $n=2,3, ...,\infty$ direct interaction bonds $b_{\rm I}$ in parallel, the summed diagrams involve arbitrary number of links, either direct or via convolution chains of $b_{\rm I}$ bonds, between the two fixed loops $\cL_i$ and $\cL_j$.}
\label{figBAM}
\end{figure}

\begin{figure}
\includegraphics[scale=1.08]{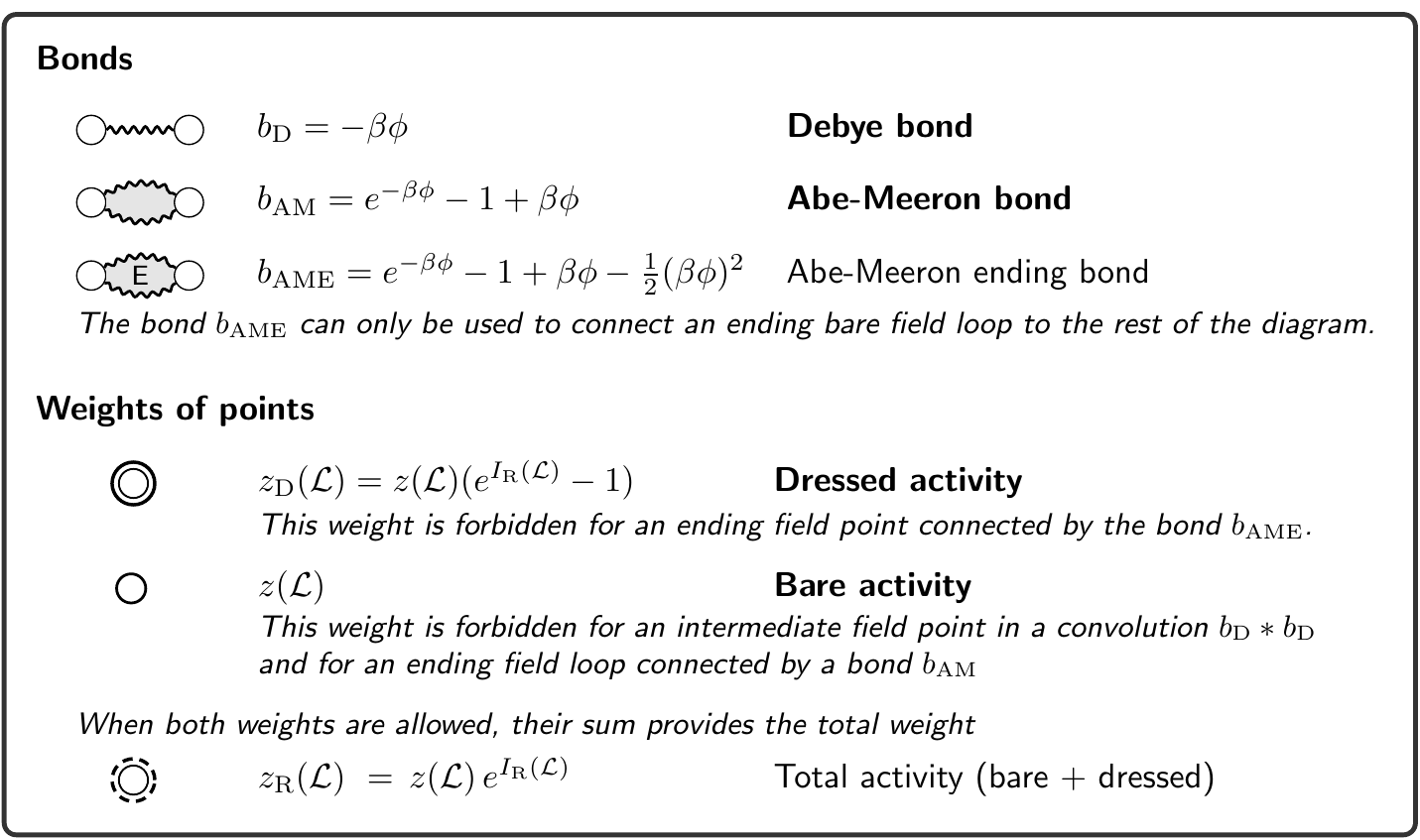}
\caption{Bonds and weights in the screened Mayer expansions for the pressure and for loop distribution functions.
In the expansion of  $\beta P$, the diagrams made with only one or two loops need a special treatment, see Eq.~\eqref{eq:ZZZ} and the comment after Eq.~\eqref{ruleVBsimplified}.
\label{FigScreenedDiagrammatics}
}
\end{figure}

The summations for all the remaining diagrams are carried out as for the one-body density~\cite{Ballenegger2002}. They generate the same screened bonds and dressed activities (see Fig.~\ref{FigScreenedDiagrammatics}).
Besides the Debye bond \eqref{DebyeBond}, the so-called Abe-Meeron bond
\be
\la{AMBond} 
b_{\rm AM}(\cL_i,\cL_j) = \exp \( b_{\rm D}(\cL_i,\cL_j) \) - 1 - b_{\rm D}(\cL_i,\cL_j) 
\ee 
is generated by summing more complex structures connecting 
the fixed pair $\cL_i$ and $\cL_j$ (see Fig.~\ref{figBAM}). 
If $\cL_i$ is an ending loop, a similar summation provides the Abe-Meeron ending bond
\be
\la{AMEBond} 
b_{\rm AME}(\cL_i,\cL_j) = \exp \( b_{\rm D}(\cL_i,\cL_j) \) - 1 - b_{\rm D}(\cL_i,\cL_j) - [b_{\rm D}(\cL_i,\cL_j)]^2/2  \; .
\ee 

\bigskip

The sum of all convolution rings involving $p \geq 1$ loops and  $(p+1)$ bonds $b_{\rm I}$ attached to a loop $\cL_i$ that is connected by more than two bonds $b_{\rm I}$ (or that is a root loop),
\be	\label{eq:27}
I_{\rm R}(\cL_i)
=\;\,
	\begin{graph}
		\twoNodesUnNamed;
				\colorNode{blanc}{1};
		\drawLinkFCC{1}{2};
	\end{graph}
\;\, + \;\,
	\begin{graph}
		\ringsAttachedMirror{3}{0.6}{1};
				\colorNode{blanc}{3};
	\end{graph}
\;\, + \;\,
	\begin{graph}
		\ringsAttached{4}{0.6}{1};
						\colorNode{blanc}{2};
	\end{graph}
\;\, + \;\,
	\begin{graph}
		\ringsAttachedMirror{5}{0.6}{1};
						\colorNode{blanc}{5};
	\end{graph}
\;\, + \;\,
\cdots 
\ee
provides the ring sum 
\begin{equation}
\la{XIIIRingSum}
I_{\rm R}(\cL_i) = \frac{1}{2}\left[ b_{\rm D}(\cL_i,\cL_i)-b_{\rm I}(\cL_i,\cL_i) \right] \;.
\end{equation}
Notice that the symmetry factor of each of these rings is $1/2$ because of the particular role of the attaching loop $\cL_i$. The sum of $n\geq 1$ such rings attached to loop $\cL_i$ generates the ring dressing factor $\left( \exp\left( I_{\rm R}(\cL_i) \right)-1 \right)$ in the definition of the dressed activity 
\begin{equation}
\la{XIIIRingActivityWeak}
z_{\rm D}(\cL_i)=z(\cL_i)\left( \exp\left( I_{\rm R}(\cL_i) \right)-1 \right)\;.
\end{equation}
%
The ring dressing factor $\exp(I_{\rm R}(\cL_i))$ accounts for the interaction energy of  loop $\cL_i$ with the surrounding polarization cloud of loops within a (non-linear) mean-field description.

\bigskip

The final screened Mayer series of the pressure reads
\begin{multline}
\la{ScreenedMayerP}
\beta P= \int D(\chi) z(\cL) + \beta P_{\rm R} + \int D(\chi) z(\cL) [e^{I_{\rm R}(\cL)}-1- I_{\rm R}(\cL)] \\
+ \sum_{{\cal P}} \frac{1}{S({\cal P})} 
\int \left[ \prod D(\cL) z^\ast(\cL) \right] 
\left[ \prod b^\ast \right]_{{\cal P}} \;
\end{multline}
where $b^*$ and $z^*$ are generic notations for the bonds and weights listed in Fig.~\ref{FigScreenedDiagrammatics}  ($\chi=\{\alpha, q,\bcX\}$).
The diagram made with a single field point, which is treated separately, provides the three first terms in this formula: the ideal term, the ring pressure \eqref{RingSum} and the contribution of a single black loop to which are attached at least two rings, represented graphically by
%
\be	\label{eq:ZZZ}
	\begin{graph}
\node[noir, thick] (1) {};
	\end{graph}
\;\quad + \;\,
	\begin{graph}
	\node[noir, thick] (1) {};
	\drawLoopNsnakeAngle{1}{90};
	\end{graph}
\;\, + \;\,
	\begin{graph}
	\node[noir, thick] (1) {};
	\drawGrayLoopSnakeAngle{1}{140}
	\drawGrayLoopSnakeAngle{1}{40}
	\end{graph}
\!\!\! + \,\ldots
\quad \equiv \quad 
	\begin{graph}
	\node[noir, thick] (1) {};
	\dressNodeSnakeGrayCircle{1};
	\end{graph}
\ee
The ring term $\beta P_{\rm R}$ reduces in the classical limit to the familiar Debye mean-field correction $\kappa^3/(12\pi)$, 
while the diagrams with two rings or more in Eq.~\eqref{eq:ZZZ} accounts for corrections 
beyond mean-field to the interaction energy of a loop with its surrounding polarization cloud~\cite{Ballenegger2017}. 
We recall that the position of an arbitrarily chosen loop in each diagram ${\cal P}$ is kept fixed thanks to translational invariance. 
For instance, in the contribution of the 
graphs \eqref{eq:ZZZ}, it is understood that the integration $D(\cL)$ is carried out over all the internal degrees of 
freedom of loop $\cL$ except its position. 

The sum in the second line is carried over all unlabelled topologically different prototype graphs ${\cal P}$ made with $N \geq 2$ black points. 
These diagrams have the same topological structure 
as the genuine Mayer diagrams. They are simply 
connected and may contain articulation points.
Each point carries a statistical weight $z^\ast(\cL)$ which is either 
\be	\label{2weights}
z(\cL)\text{\quad(bare loop)}\qquad \text{or}\qquad z_{\rm D}(\cL)
\text{\quad(dressed loop)}\;.
\ee
When both weights are allowed for a point, the two possibilities can be added together to form the weight
\be
\la{RingActivity}
z_{\rm R}(\cL) = z(\cL) + z_{\rm D}(\cL) = z(\cL) \exp(I_{\rm R}(\cL))\;.
\ee
There exists three possible bonds $b^* = b_{\rm D}$, $b_{\rm AM}$ and $b_{\rm AME}$. The bond $b_{\rm AME}$ can only be used to connect an ending bare field loop to the rest of the diagram  (this rest can consist in a single bare or dressed loop in the particular case of a diagram made with two loops).
%
In general, the two weights~\eqref{2weights} are possible, except in the following cases~:
\begin{itemize}
\item If $\cL$ is an ending field loop connected by a bond $b_{\rm AM}$ or $b_{\rm AME}$, its weight is
\be
\label{ruleVBsimplified}
z^*(\cL) =
\begin{cases}
z(\cL)& \text{if $b^* = b_{\rm AME}$} \\
 z_{\rm D}(\cL) & \text{if $b^* = b_{\rm AM}$} \\
\end{cases}
\ee
\item If $\cL$ is an intermediate field loop in a convolution $b_{\rm D}*b_{\rm D}$ of two Debye bonds, its weight is $z_{\rm D}(\cL)$.
\end{itemize}
Moreover, the case of a diagram made with only two loops is special, because both loops are then ending. The case $b^*=b_{\rm AME}$ in rule \eqref{ruleVBsimplified} is then modified to allow not only the case where both field loops are bare, but also the case where one loop is bare and one loop is dressed (see Fig.~6).

\bigskip

\begin{figure}[t]
\includegraphics[scale=1.0]{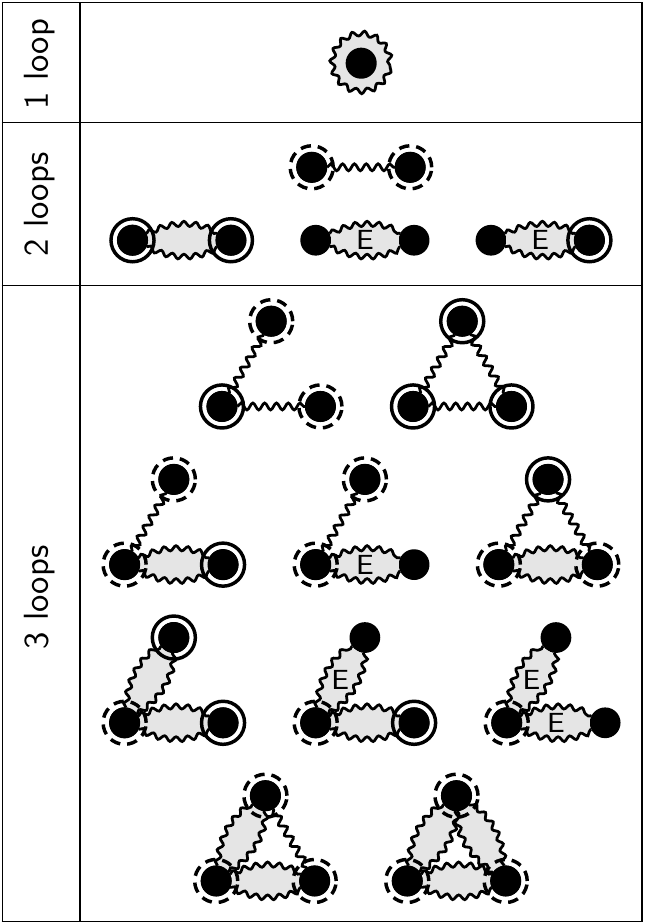}
\caption{All diagrams involving one, two or three loops in the screened Mayer series of the pressure.}
\label{figDiagrams123}
\end{figure}

These diagrammatic ingredients and rules are summarized in Fig.~\ref{FigScreenedDiagrammatics}. These rules are valid not only for diagrams with $N \geq 2$ points in the screened Mayer series of the pressure, but also for all diagrams in the screened Mayer series of any loop distribution function $\rho^{(n)}(\cL_1, ..., \cL_n)$. In the latter case, each diagram contains $n$ root points and an arbitrary number of field points. Fig.~\ref{figDiagrams123} shows all diagrams in the screened Mayer series of the pressure made with 1, 2 or 3 loops.
%

\bigskip

The central quantity is the Debye bond 
$b_{\rm D}(\cL_i,\cL_j)=-\beta e_{\alpha_i} e_{\alpha_j} \phi(\cL_i,\cL_j)$. As shown in Ref.~\cite{Ballenegger2002}, 
$\phi$ decays as $1/r^3$ at large distances $r$ 
between two loops. Thus bonds $b_{\rm AM}$ and $b_{\rm AME}$ decay respectively as $1/r^6$ and $1/r^9$, and they are integrable. 
The bond $b_{\rm D}$ decays as $\phi$ itself, \textit{i.e.} as $1/r^3$, which is at the border line for integrability. 
Accordingly, the graphs with ending loops connected to the rest of the diagram by bonds $b_{\rm D}$ have to be dealt with some care. 
In fact, since the corresponding weight of the ending loop, $z_{\rm R}(\cL)$, is an even function of the loop shape $\bcX(s)$, if we proceed first to
functional integrations over the shape, then the $1/r^3$-algebraic tails vanish, because their amplitudes are odd functions of $\bcX(s)$ and
every prototype graph provides a finite contribution~\cite{Ballenegger2002}.

\subsection{Link with the activity-series for the particle densities}
\label{SectionLink}

The screened activity expansion of the loop density can be readily inferred from 
the expansion~(\ref{ScreenedMayerP}) of the pressure by using 
\be
\la{FunctionalDerivativeP}
\rho(\cL_{\rm a})=z(\cL_{\rm a}) \frac{ \delta \beta P}{\delta z(\cL_{\rm a})} \;.
\ee
The activity-expansion of the particle densities follows then from 
\be
\la{ParticleDensityLoop}
\rho_{\alpha_{\rm a}} = \sum_{q_{\rm a}=1}^{\infty}q_{\rm a} \;\int D_{q}(\bcX_{\rm a}) \; \rho(\cL_{\rm a}) \; .
\ee
Notice that $P=k_{\rm B}T \ln{\Xi}/\Lambda$ is viewed in Eq.~\eqref{FunctionalDerivativeP} as a functional of the loop activity $z(\cL)$, which is present in Eq.~\eqref{IX.QMG6} and also in bonds and weights of the resummed diagrammatics. We consider here that the function $z(\cL)$ can also vary with the root position $\pmb{x}$ of the loop $\cL$, as it does in an inhomogeneous system. The rules of functional derivatives generate then straightforwardly the expressions for the potentially space-dependent loop density $\rho(\cL)$.
When computing the particle density $\rho_\alpha$ in a homogeneous plasma, as in sections \ref{LDExpansionEOS} and \ref{DDR}, the functional derivative can be replaced by an ordinary partial derivative with respect to the one-dimensional variable $z_\alpha$.

\bigskip

The functional derivative of each prototype diagram ${\cal P}$ is calculated by either whitening a black loop $\cL$
with weight $z(\cL)$ into the root loop $\cL_{\rm a}$ with weight $z(\cL_{\rm a})$ or by taking the functional derivative with 
respect to $z(\cL_{\rm a})$ of $\beta P_{\rm R}$ and $b_{\rm D}(\cL_i,\cL_j)$, namely
\be
\la{FDRingP}
z(\cL_{\rm a}) \frac{ \delta \beta P_{\rm R}}{\delta z(\cL_{\rm a})}= z(\cL_{\rm a}) I_{\rm R}(\cL_{\rm a})
\ee 
and 
\be
\la{FDphi}
z(\cL_{\rm a}) \frac{ \delta b_{\rm D}(\cL_i,\cL_j)}{\delta z(\cL_{\rm a})} = z(\cL_{\rm a})b_{\rm D}(\cL_i,\cL_{\rm a})b_{\rm D}(\cL_{\rm a},\cL_j) \; .
\ee
Note that the functional derivatives of the dressed activities and of the other bonds, which can be all expressed in terms of 
$b_{\rm D}$, are then obtained by using Eq.~(\ref{FDphi}).
In particular, the derivative of the ring factor $I_{\rm R}(\cL_i)$ generates the bond $\frac{1}{2} b_{\rm D}^2(\cL_a, \cL_i)$.
This calculation provides 
\be
\la{IX.QMG88}
\rho(\cL_{\rm a}) =   \sum_{{\cal P}_{\rm a}} \frac{1}{S({\cal P}_{\rm a})} 
\int \left[ \prod D(\cL) z^\ast(\cL) \right] 
\left[ \prod b^\ast \right]_{{\cal P}_{\rm a}} \; . 
\end{equation}
which can be also obtained by a direct Abe-Meeron summation of the Mayer diagrammatic series for the loop density~\cite{Ballenegger2002}.  In obtaining Eq.~\eqref{IX.QMG88}, we have used that each bond $\frac{1}{2} b_{\rm D}^2(\cL_a, \cL_i)$ can be added to the bond $b_{\rm AME}(\cL_a, \cL_i)$ in diagrams with the same topological structure to provide the bond $b_{\rm AM}$.

\bigskip

The  prototype diagrams ${\cal P}_{\rm a}$  have 
one root (white) point with weight $z_{\rm R}(\cL_a)$, $N=0, 1, 2, ...$ field (black) points and obey the diagrammatical rules summarized in Fig.~\ref{FigScreenedDiagrammatics}. The first few diagrams in the series~\eqref{IX.QMG88} are
\be
\rho(\cL) = 
	\begin{graph}
		\node[blanc] (1) {};
		\dressNodeDashedCircle{1};
	\end{graph}
	\;+\;
	\begin{graph}
		\twoNodesUnNamed;
		\colorNode{blanc}{1};
		\dressNodeDashedCircle{1};
		\dressNodeDashedCircle{2};
		\drawLinkD{1}{2};
	\end{graph}
	\;+\;
	\begin{graph}
		\twoNodesUnNamed;
		\drawLinkAME{1}{2};
		\colorNode{blanc}{1};
		\dressNodeDashedCircle{1};
	\end{graph}
	\;+\;
	\begin{graph}
		\twoNodesUnNamed;
		\drawLinkAM{1}{2};
		\colorNode{blanc}{1};
		\dressNodeDashedCircle{1};
		\dressNodeCircle{2}; 
	\end{graph}
	\;+\; \ldots
\ee
It can be checked that there are 16 topologically different diagrams made with 3 loops.

\subsection{Neutrality and low-density expansion of the EOS}
\label{Section3.D}

\subsubsection{Neutrality, pseudo-neutrality and Debye dressing}
\la{ChargeNeutralityDensity}

The collective electrostatic effects in a finite box 
which enforce charge neutrality in the grand-canonical ensemble (see Section~\ref{sec:S2}), do not show in each individual term of the activity series, 
where only a finite number of particles intervene. This is particularly striking for the ideal contribution 
in series~(\ref{IX.QMG88}) for the particle density, whose Maxwell-Boltzmann 
(weak-degeneracy) limit involves a single particle. The pseudo-neutrality condition, 
$
\sum_\alpha e_\alpha z_\alpha=0
$, 
which can be safely imposed as argued in 
Section~\ref{sec:S2}, restores the previous collective electrostatic effects at this lowest order in the particle activities
Such effects might otherwise be erased in the series~(\ref{ScreenedMayerP}) and~\eqref{IX.QMG88} since the boundaries have been sent to infinity 
without worrying about surface effects.
Importantly, the pseudo-neutrality condition ensures moreover that, at a given order in the small activities~$z$, 
the expansions~(\ref{ScreenedMayerP}) and~(\ref{IX.QMG88}) for the pressure and the particle densities can be calculated 
by keeping only a finite number of diagrams. We show first this point, and demonstrate then that local charge neutrality is always ensured due to the structure of these series.

\bigskip

Let us consider the series~(\ref{ScreenedMayerP}) for the pressure.
For any given graph ${\cal P}$, there exists a Debye dressed graph ${\cal P}^{\rm D}$ obtained by 
adding a black loop $\cL$ with weight $z(\cL)$ connected to ${\cal P}$ \textsl{via} a single bond $b_{\rm D}(\cL,\cL')$ 
where $\cL'$ is a black loop inside ${\cal P}$, that is
\be
\label{diagramPD}
{\cal P}^{\rm D} = 
	\begin{graph}
		\draw[black,fill=gray] (0,0cm) circle [radius=0.75 cm];
		\node[transparent] (3) at (+0.2,0.45) [label=below:${\cal P}$] {};
		\node[noir] (1) at (-0.35,0) [label={[xshift=1mm,yshift=0.75mm]below:${\cal L}'$}] {};
		\node[noir] (2) at (-1.75,0) [label={[xshift=1mm,yshift=0.75mm]below:${\cal L}$}] {};
		\drawLinkD{1}{2};
	\end{graph}
\ee
In the low-activity limit, the 
potential $\phi$ reduces to its classical Debye counterpart~\cite{Ballenegger2002}, so 
\be
\la{DebyeBondLowA}
b_{\rm D}(\cL,\cL') \sim -\beta q_\alpha e_\alpha q_{\alpha'}e_{\alpha'} \frac{\exp(-\kappa_z |\bx-\bx'|  )}{|\bx-\bx'|}
\ee
and where we have used
\be
\la{KappaLowA}
\kappa^2(0,0) \sim \kappa_z^2= 4 \pi \beta \sum_\gamma  e_\gamma^2 z_\gamma \; .
\ee
At leading order in the small activities, the contribution of the graph ${\cal P}^{\rm D}$ is obtained by keeping only the loop 
$\cL$ made with a single particle, i.e. $q_\alpha=1$, while the bond $b_{\rm D}(\cL,\cL')$ is replaced by its classical Debye 
expression~(\ref{DebyeBondLowA}). The leading contribution of ${\cal P}^{\rm D}$ reduces hence to that of graph ${\cal P}$ multiplied
by 
\begin{multline}
\la{DebyeBondLowAbis}
\int D(\cL) z(\cL) \; b_{\rm D}(\cL,\cL') \sim -\beta  q_{\alpha'}e_{\alpha'} \sum_\alpha e_\alpha z_\alpha 
\int \d \bx \frac{\exp(-\kappa_z |\bx-\bx'|  )}{|\bx-\bx'|} \\
=-\frac{4 \pi \beta  q_{\alpha'}e_{\alpha'}}{\kappa_z^2} \sum_\alpha e_\alpha z_\alpha \; .
\end{multline}
At leading order, the contribution of ${\cal P}^{\rm D}$ has obviously the same order as 
that of ${\cal P}$ for arbitrary sets  $\{ z_\alpha\}$ of particle activities. In other words, in order to compute the pressure at 
a given order for such sets, one would have to keep an infinite number of graphs in the series~(\ref{ScreenedMayerP}), since the dressing 
of a given ${\cal P}$  can be repeated an arbitrary number of times. This infinite sum might actually not converge, meaning 
that the screened Mayer series~\eqref{ScreenedMayerP} and~\eqref{IX.QMG88} in an unbounded volume might make sense only 
when the pseudo-neutrality condition is imposed. The pseudo-neutrality condition~(\ref{PseudoNeutrality}) greatly simplifies the calculations at a given order. Indeed, the graph ${\cal P}^{\rm D}$ contributes then at a higher order than graph ${\cal P}$. Only a finite number of graphs ${\cal P}$ in the series~(\ref{ScreenedMayerP}) and \eqref{IX.QMG88} need then to be kept.

\bigskip

The property of a quantum plasma to be locally charge neutral at equilibrium 
can be proved by combining Eq.~\eqref{ParticleDensityLoop} for the particle density with the Mayer series \eqref{IX.QMG88} for the density of loops. This proof is based on a simple Debye-dressing mechanism at work in the resulting series for the particle densities.

\begin{figure}[h]
\includegraphics[scale=1.15]{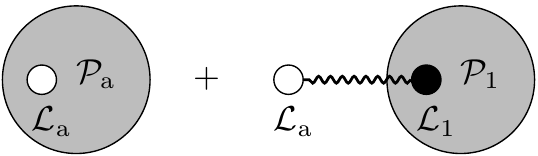}
\caption{
A diagram ${\cal P}_{\rm a} \in {\cal C}_{\rm a}$, and its Debye-dressed companion diagram ${\cal P}_{\rm a}^{\rm D}$.  In diagram ${\cal P}_{\rm a}$, loop ${\cal L}_{\rm a}$ cannot be a bare loop connected only by a bond $b_{\rm D}$, whereas ${\cal L}_{\rm a}$ is precisely such a loop in diagram ${\cal P}^{\rm D}_{\rm a}$.
}
\label{figDebyeDressingDensity}
\end{figure}
%

Remembering that each loop in a prototype diagram is either bare or dressed, we classify the diagrams into two groups: the class ${\cal C}^{\rm D}_{\rm a}$ of diagrams where the root loop $\cL_a$ is an ending bare loop connected only by a Debye bond $b_{\rm D}$, and the class ${\cal C}_{\rm a}$ containing all other diagrams.
For any diagram ${\cal P}_{\rm a}$ in the class ${\cal C}_{\rm a}$, including the most simple diagram made with only one single bare loop, there exists a unique corresponding diagram ${\cal P}^{\rm D}_{\rm a}$ in the class ${\cal C}^{\rm D}_{\rm a}$ where $\cL_{\rm a}$ is a bare loop connected by a (single) bond $b_{\rm D}$ to a subdiagram identical to ${\cal P}_{\rm a}$ but where the root point is replaced by a field point, which we label $\cL_1$ 
(see Fig.~\ref{figDebyeDressingDensity}). This establishes a one-to-one correspondence between the diagrams in the two classes because no convolution 
$b_{\rm D}*b_{\rm D}$ with an intermediate bare field loop is allowed in the prototype diagrams. The diagram ${\cal P}^{\rm D}_{{\rm a}}$ is said to be the Debye-dressed companion of diagram ${\cal P}_{\rm a}$.

The contribution of diagram ${\cal P}^{\rm D}_{\rm a}$ to the density $\rho_{\alpha_{\rm a}}$,
\be
\la{GraphDDensityBareBis}
\rho_{\alpha_{\rm a}}[{\cal P}^{\rm D}_{\rm a}]=- \frac{4 \pi \beta e_{\alpha_{\rm a}}}{\kappa^2(0,0) } 
\sum_{q_{\rm a}=1}^\infty q_{\rm a}^2 \int D_{q_{\rm a}}(\bcX_{\rm a}(\cdot)) z(\cL_{\rm a})
\sum_{\alpha_{1}} e_{\alpha_{1}} \rho_{\alpha_{1}}[{\cal P}_{1}] \; ,
\ee
is easily calculated with the frequency decomposition~(\ref{IX.QMG48}) 
of $\phi(\cL_{\rm a},\cL_1)$ and translation invariance. 
Notice that this contribution has the same order, when $z\to 0$, as the one of diagram $\rho_{\alpha_{\rm a}}[{\cal P}_{\rm a}]$ because $z(\cL) \propto z$ and $\kappa^2(0,0)\propto z$.
Using expression~(\ref{IX.QMG45a}) for $\kappa^2(0,0)$, this contribution to the local charge density reduces to
\be
\la{GraphDChargeDensityBare}
\sum_{\alpha_{\rm a}}  e_{\alpha_{\rm a}}\rho_{\alpha_{\rm a}}[{\cal P}^{\rm D}_{\rm a}]= 
- \sum_{\alpha_{\rm a}}  e_{\alpha_{\rm a}}\rho_{\alpha_{\rm a}}[{\cal P}_{\rm a}]\; . 
\ee
The contribution to the local charge density of any diagram ${\cal P}_{\rm a}$ is thus exactly compensated by the contribution of its 
companion diagram ${\cal P}^{\rm D}_{\rm a}$. The local charge density therefore vanishes. Notice that this proof does not require the pseudo-neutrality condition to be satisfied. If pseudo-neutrality does not hold, there is an infinite number of diagrams contributing at the same order, rendering the proof only formal, whereas there is only a finite number of diagrams contributing at a given order when the pseudo-neutrality condition holds. 

\subsubsection{Expansion of the EOS at order $\rho^2$}
\la{LDExpansionEOS}

The low-density expansion of the EOS has been computed up to order $\rho^{5/2}$ 
by various methods~\cite{Ebeling1967,Kraeft1986,Alastuey1992,deWitt1995,Brown2001}, 
which all provide eventually identical physical predictions~\cite{Alastuey2015}. Our purpose in this section is to illustrate the efficiency of the method 
based on the screened activity expansion~\eqref{ScreenedMayerP} of the pressure by outlining how all terms up to order $\rho^2$ in the EOS can be computed. 

\bigskip

\mbox{}
\vskip -6mm
Since at low densities, $z \sim \rho$, in order to obtain the EOS at order $\rho^2$, we need to start with the $z$-expansion of $\beta P$ at the order 
$z^2$. We assume that the pseudo-neutrality condition holds. Then, at this order, one only needs to consider the diagrams made with 1 or 2 loops, i.e. the first five diagrams in Fig.~\ref{figDiagrams123}. In the first diagram made with 2 loops in this figure, we can discard the cases where one or both loops are bare because their contributions are $o(z^2)$ thanks to pseudo-neutrality.
Since $z_{\rm D} \propto z^{3/2}$, the next diagram and the last one made with two loops are of order $z^{5/2}$ and can hence also be discarded.
Only three diagrams remain,
\be
\label{P2}
\beta P \quad=\quad
	\begin{graph}
	\node[noir, thick] (1) {};
	\dressNodeSnakeGrayCircle{1};
	\end{graph}
\quad+\quad
	\begin{graph}
		\twoNodesUnNamed;
		\drawLinkD{1}{2};
		\dressNodeCircle{1};
		\dressNodeCircle{2};
	\end{graph}
\quad+\quad
	\begin{graph}
		\twoNodesUnNamed;
		\drawLinkAME{1}{2};
	\end{graph} \quad+\; o(z^2)\;.
\ee


\noindent 
Significantly more diagrams contribute to the density at the same order,
\begin{align}	\notag
\rho_\alpha \quad=\quad
	\begin{graph}
		\node[blanc] (1) {};
		\dressNodeDashedCircle{1};
	\end{graph}
	\;&+\;
	\begin{graph}
		\twoNodesUnNamed;
		\colorNode{blanc}{1};
		\dressNodeCircle{1};
		\dressNodeCircle{2};
		\drawLinkD{1}{2};
	\end{graph}
	\;+\;
	\begin{graph}
		\twoNodesUnNamed;
		\drawLinkAME{1}{2};
		\colorNode{blanc}{1};
	\end{graph}
	\;+\;
	\begin{graph}
		\twoNodesUnNamed;
		\drawLinkAM{1}{2};
		\colorNode{blanc}{1};
		\dressNodeCircle{2}; 
	\end{graph} \\ \notag
	&+\;
	\begin{graph}
		\node[blanc] (1) {};
		\node[noir] (2) at($(1) + (0:1) $){};
		\node[noir] (3) at($(1) + (60:1) $){};
		\drawLinkD{1}{2};
		\drawLinkD{1}{3};
		\colorNode{blanc}{1};
		\dressNodeCircle{2};
		\dressNodeCircle{3};
	\end{graph}
	\;+\;
	\begin{graph}
		\node[blanc] (1) {};
		\node[noir] (2) at($(1) + (0:1) $){};
		\node[noir] (3) at($(1) + (60:1) $){};
		\drawLinkAM{1}{2};
		\drawLinkD{2}{3};
		\colorNode{blanc}{1};
		\dressNodeCircle{3};
	\end{graph}
	\;+\;
	\begin{graph}
		\node[blanc] (1) {};
		\node[noir] (2) at($(1) + (0:1) $){};
		\node[noir] (3) at($(1) + (60:1) $){};
		\drawLinkD{1}{2};
		\drawLinkD{1}{3};
		\drawLinkAM{2}{3}
		\colorNode{blanc}{1};
	\end{graph}	\\
	&+\,\text{7 Debye-dressed companion diagrams}
	\quad+\;o(z^2).
\label{DensityPartialDerivativeP}
\end{align}
Since Eq.~\eqref{DensityPartialDerivativeP} includes all companion DD diagrams, it leads to particle densities that satisfy the charge neutrality 
$\sum_\alpha e_\alpha \rho_\alpha = 0$, as shown in the previous section. 
In the last four drawn diagrams, the dressed weight $z_{\rm D}$ of the root point has been discarded into the $o(z^2)$ remainder.

The seven DD companion diagrams in Eq.~\eqref{DensityPartialDerivativeP} are
\begin{multline}
\la{classicalDD}
	\begin{graph}
		\node[blanc,scale=0.7,semithick] (1) {} ;
		\node[noir] [right of=1] (2) {};
		\dressNodeDashedCircle{2};
		\drawLinkDclas{1}{2}
	\end{graph}
	\;\;+\;\;
	\begin{graph}
		\node[blanc,scale=0.7,semithick] (1) {} ;
		\node[noir] [right of=1] (2) {};
		\node[noir] (3) at($(1) + (60:1) $){};
		\dressNodeCircle{2};
		\dressNodeCircle{3};
		\drawLinkDclas{1}{2}
		\drawLinkD{2}{3};
	\end{graph}
	\;\;+\;\;
	\begin{graph}
		\node[blanc,scale=0.7,semithick] (1) {} ;
		\node[noir] [right of=1] (2) {};
		\node[noir] (3) at($(1) + (60:1) $){};
		\drawLinkDclas{1}{2}
		\drawLinkAME{2}{3};
	\end{graph}
	\;\;\;\;\\
	+\;
	\begin{graph}
		\node[blanc,scale=0.7,semithick] (1) {} ;
		\node[noir] [right of=1] (2) {};
		\node[noir] (3) at($(1) + (60:1) $){};
		\drawLinkDclas{1}{2}
		\drawLinkAM{2}{3};
		\dressNodeCircle{3};
		\dressNodeCircle{2}; 
	\end{graph}
	\;\;+\;\;
		\begin{graph}
		\node[blanc,scale=0.7,semithick] (1) {} ;
		\node[noir] [right of=1] (2) {};
		\node[noir] [above of=1,yshift=-1mm] (3) {};
		\node[noir] [above of=2,yshift=-1mm] (4) {};
		\drawLinkDclas{1}{2}
		\drawLinkD{3}{4};
		\drawLinkAM{2}{4};
		\dressNodeCircle{3};
		\dressNodeCircle{4};
	\end{graph}
		\;\;+\;\;
	\begin{graph}
		\node[blanc,scale=0.7,semithick] (1) {} ;
		\node[noir] [right of=1] (2) {};
		\node[noir] [above of=1,yshift=-1mm] (3) {};
		\node[noir] [above of=2,yshift=-1mm] (4) {};
		\drawLinkDclas{1}{2}
		\drawLinkD{2}{3};
		\drawLinkD{2}{4};
		\dressNodeCircle{3};
		\dressNodeCircle{4};
	\end{graph}
	\;\;+\;\;
		\begin{graph}
		\node[blanc,scale=0.7,semithick] (1) {} ;
		\node[noir] [right of=1] (2) {};
		\node[noir] [above of=1,yshift=-1mm] (3) {};
		\node[noir] [above of=2,yshift=-1mm] (4) {};
		\drawLinkDclas{1}{2}
		\drawLinkD{2}{3}
		\drawLinkD{2}{4}
		\drawLinkD{3}{4};
		\drawLinkAM{3}{4};
	\end{graph}
\end{multline}
At the considered order $O(z^2)$, the root point  is a single particule of species $\alpha$, i.e. a loop with $q=1$, with weight $z_{\alpha}$, and the wavy and dotted bond represents the classical Debye screened bond given by Eq.~\eqref{IX.QMG47} where $\psi_{\rm loop}(\pmb{r},s)$ is replaced by $\exp(-\kappa_z r)/r$ with 
$r = |\pmb{r}|$ while $\kappa^2_z$ is defined in Eq.~\eqref{KappaLowA}. The contribution of a diagram ${\cal P}^{\rm D}$ companion of ${\cal P}$
to density $\rho_{\alpha}$, is given by the general 
formula~(\ref{GraphDDensityBareBis}) with $\alpha_{\rm a}=\alpha$ , which then reduces to 
\be
\la{CompanionClassical}
\rho_{\alpha}[{\cal P}^{\rm D}]=- \frac{4 \pi \beta e_{\alpha} z_{\alpha}}{\kappa_z^2 } 
\sum_{\gamma} e_{\gamma} \rho_{\gamma}[{\cal P}]
\ee
discarding terms of order $o(z^2)$. Since the contributions $\rho_{\gamma}[{\cal P}]$ of the 7 diagrams~(\ref{DensityPartialDerivativeP}) are obtained by 
merely taking the partial derivative $ z_\alpha \partial(\beta P)/\partial z_\alpha$ of the retained pressure diagrams 
at the same order $O(z^2)$, we see that the contributions $\rho_{\alpha}[{\cal P}^{\rm D}]$ of their companions at the same order $O(z^2)$
are also fully determined by the pressure diagrams~(\ref{P2}). 

\bigskip

After computing the pressure at order $z^2$, denoted $P^{(2)}$ and the density at the same order, 
denoted $\rho_{\alpha}^{(2)}(\{z_\gamma \})$, one determines the equation of state $P(T,\{\rho_\gamma\})$.
The ${\cal S}$ pseudo-neutral activities $\{z_\gamma\}$ up to order $\rho^2$ included, denoted $\{z_\gamma^{(2)}\}$, are 
first obtained as function of the physical 
densities $\{\rho_\alpha\}$ by inverting perturbatively the ${\cal S}$ 
independent relations 
\begin{align}
\la{DensityActivity}
&\rho_{\alpha}^{(2)}(\{z_\gamma^{(2)} \}) = \rho_{\alpha}, 
\quad  \quad \alpha_{\rm a}=1,...,{\cal S}-1  \nonumber \\
& \sum_\gamma e_\gamma z_\gamma^{(2)}= 0 \; .
\end{align}
The required EOS up at order $\rho^2$ included follows from inserting $\{z_\gamma^{(2)}\}$ 
into Eq.~\eqref{P2}, namely
\be
\la{EOS2}
\beta P(\beta;\{\rho_\alpha \})=P^{(2)}(\beta;\{z_\gamma^{(2)}(\{\rho_\alpha \}) \}) + o(\rho^2) \; .
\ee
It can be checked that the known density expansion of the pressure is indeed recovered. 

\bigskip

This calculation illustrates the usefulness of the activity series for the pressure which considerably reduces the number of diagrams 
which need to be computed. A similar scheme can be repeated at the next orders. 
However, even if the number of diagrams which need to be computed is reduced with respect to other methods, it remains a formidable task to 
obtain the terms of order $\rho^3$. In particular, one has to take into account 
quantum effects embedded in $\phi$ and $I_{\rm R}$, so a classical treatment of the dressing mechanism is no longer sufficient.

\subsubsection{Computation of the densities by differentiation of the pressure with Debye-dressed activities}
\label{sectionDDActivities}

It is instructive to interpret the previous results in terms of Debye-Dressed activities.
Firstly, we note that the $7$  companion-diagrams~(\ref{classicalDD}) in the density series arise from other diagrams in the pressure series 
which do not contribute at order $z^2$ by virtue of the pseudo-neutrality condition.  In fact, the pseudo-neutrality condition must be applied only after the derivative $ z_\alpha \partial(\beta P)/\partial z_\alpha$ has been taken, whereas it has already been applied in an expression like~\eqref{P2}. Since a contribution to the pressure that vanishes by pseudo-neutrality can have a non-vanishing derivative with respect to $z$, and 
hence a non-vanishing contribution to the density $\rho_\alpha$, one needs to consider also such contributions in the pressure series. 
These contributions can be seen as decorations of the diagrams that do not increase their order. The Debye-Dressing (DD) of a loop in a diagram is an example of such a contribution (recall Eq.~\eqref{diagramPD}).
Adding DD decorations successively to each points in the three diagrams of Eq.~\eqref{P2} provides a set of diagrams which generate, after differentiation, 
the corresponding density diagrams in Eq.~\eqref{DensityPartialDerivativeP} except for the last four companion diagrams in Eq.~\eqref{classicalDD}.
In fact, these four diagrams arise from other diagrams in the pressure series which are nothing but the same diagrams where the root white point is transformed 
into a black point. Nevertheless, we will show that all the 14 diagrams can be computed by direct partial differentiation of only the 3 pressure diagrams~(\ref{P2}) when Debye-dressed activities are used.

\bigskip 

Let us define the Debye-Dressed activity
\be
\la{DDfunction}
z_{\alpha}^{\rm DD} =z_\alpha \left( 1 -\frac{4 \pi \beta e_\alpha }{\kappa_z^2} \sum_\gamma e_\gamma z_{\gamma} \right) \; ,
\ee
where the second term in Eq.~\eqref{DDfunction} is the classical DD factor~\eqref{DebyeBondLowAbis}. 
Let us replace, in the 3 pressure diagrams~(\ref{P2}), 
the weights $z_\alpha$ of the black points by $z_{\alpha}^{\rm DD}$.
This amounts to decorate the considered diagrams and it provides after differentiation 
the first 10 diagrams in Eq.~\eqref{DensityPartialDerivativeP}. Since the remaining four companion diagrams in Eq.~\eqref{classicalDD} are 
associated with diagrams obtained by taking the derivative of the ring factor $I_{\rm R}$ and of the bond $b_{\rm D}$, we see that 
if we also replace $z_\alpha$ by $z_{\alpha}^{\rm DD}$ in both $I_{\rm R}$ and  $b_{\rm D}$ in the 3 pressure diagrams~(\ref{P2}), the standard rules of partial derivatives of composite functions generate the last four companion diagrams in Eq.~\eqref{classicalDD}, thanks 
to the expression~(\ref{PartialDerivativeDDbis}) of the partial derivative $\partial z_{\theta}^{\rm DD}/\partial z_{\alpha}$. Again, and 
as mentioned above, this partial derivative 
has to be calculated for any set of independent activities, while the pseudo-neutral condition is applied afterward. 
Hence, if we replace all the activities $z_\alpha$ by the functions $z_{\alpha}^{\rm DD}$ in the weights and bonds of the 3 pressure 
diagrams~(\ref{P2}), the corresponding function $P^{\rm DD}$ is such that the derivative $z_\alpha \partial(\beta P^{\rm DD})/\partial z_\alpha$ generates automatically the values of all the 14 diagrams in Eq.~\eqref{DensityPartialDerivativeP} at order $z^2$ included.

\section{Debye-Dressing neutralization prescription}
\la{DDR}
 
Let us consider a given approximation $P_{\rm A}(\beta;\{ z_i \})$ obtained by selecting specific diagrams in  
the series~(\ref{ScreenedMayerP}). As discussed in Section~\ref{sec:S23} for any approximation, the densities inferred from 
$P_{\rm A}(\beta;\{ z_i \})$ \textsl{via} the standard identities~(\ref{DensityPressureA}) do not necessarily satisfy the local charge neutrality. 
We introduced a general prescription, based on the Neutral-Group activities, which systematically circumvents this drawback. Here, 
we propose a different, but closely related, general method inspired by the Debye-Dressing mechanism described and applied to the first terms of 
the pressure and density series up to order $z^2$.

\subsection{Debye-Dressing neutralization prescription}
\la{Sec:S41}

The Debye-dressed diagrams (see Fig.~\ref{figDebyeDressingDensity}) in the series for the particle densities are crucial for ensuring the local charge neutrality. If a given diagram contributes to the particle densities in a way that breaks the local charge neutrality, adding the contribution of its DD companion diagram is sufficient to restore electro-neutrality. Inspired by this simple mechanism, one can define the following heuristic Debye-Dressing neutralization prescription
\be
\la{DressingRecipe}
\rho_{\alpha} = z_{\alpha} \frac{\partial P_{\rm A}}{\partial z_{\alpha} }
-\frac{4 \pi \beta e_\alpha z_{\alpha}}{\kappa_z^2} \sum_\gamma e_\gamma z_{\gamma} \frac{\partial P_{\rm A} }{\partial z_{\gamma} } \; .
\ee 
The two terms in this equation are the analogs of the two diagrams in Fig.~\ref{figDebyeDressingDensity}. The classical expression for the Debye-dressing factor is used in Eq.~\eqref{DressingRecipe}, as in Section~(\ref{Section3.D}) for exact calculations at order $z^2$, because it is sufficient to ensure electroneutrality.
%
%
We stress that the partial derivatives in the dressed expression~(\ref{DressingRecipe}) are calculated as usual, namely for independent activities $z_\alpha$.
However, at the end, their values are determined for a set $\{ z_{\alpha} \} $ satisfying the pseudo-neutrality 
condition~(\ref{PseudoNeutrality}).   
The local charge neutrality is automatically satisfied by the dressed densities~(\ref{DressingRecipe}). 
If the undressed densities
\be
\la{UndressedD}
z_{\alpha} \frac{\partial P_{\rm A}}{\partial z_{\alpha} }
\ee 
carry a non-zero net charge $q_{\rm exc}$, the dressed density $\rho_\alpha$~(\ref{DressingRecipe}) is shifted from its undressed counterpart 
by a term proportional to  $e_\alpha z_\alpha q_{\rm exc}$. As it should, this shift vanishes if $q_{\rm exc}=0$, 
namely if the undressed densities~(\ref{UndressedD})
already satisfy local charge neutrality. 

\bigskip 


In Section~\ref{sectionDDActivities}, Debye-dressed activities $\{z^{\rm DD}_\alpha\}$ have been introduced to take into account systematically, at any order in the particle activities, the classical Debye screening effect when determining the particle densities associated with some diagrams 
in the pressure series~\eqref{ScreenedMayerP}. The DD activities can therefore also be used to determine, from an 
approximate expression~$P_{\rm A}(\beta;\{ z_i \})$ for the pressure, particle densities that satisfy electroneutrality, and also a grand-potential 
from which these densities derive. 
Let us show that this way of ensuring electroneutrality, which is exact at order~$z^2$, leads to the same particle 
densities as the prescription~\eqref{DressingRecipe}.
%
%
In that approach, to any approximation $P_{\rm A}(\beta;\{ z_i \})$ for the pressure, we introduce the associated approximation  
\be
\la{DDpressure}
P_{\rm A}^{\rm DD}(\beta;\{ z_\alpha \})=P_{\rm A}(\beta;\{ z_\alpha^{\rm DD} \})
\ee
where each $z_\alpha$ in $P_{\rm A}$ is replaced by the Debye-Dressed function~\eqref{DDfunction} of the activities. The particle densities inferred 
from $P_{\rm A}^{\rm DD}$, namely
\be
\la{DensityPressureADD}
\rho_{\alpha} = z_{\alpha} \frac{\partial P_{\rm A}^{\rm DD}}{\partial z_{\alpha}}(T;\{ z_{\gamma} \}) \; ,
\ee
can be calculated by applying the rules of composition of partial derivatives, 
\begin{equation}
\la{DensityPressureADDbis}
\rho_{\alpha} = z_{\alpha} \frac{\partial }{\partial z_{\alpha}} P_{\rm A}(T;\{ z_\gamma^{\rm DD} \}) \nonumber \\
=  \sum_{\theta=1}^{\cS}  \frac{\partial P_{\rm A}}{\partial z_{\theta}} (T;\{z_\gamma^{\rm DD} \}) 
z_{\alpha} \frac{\partial z_{\theta}^{\rm DD}}{\partial z_{\alpha}}\; .
\end{equation}
The partial derivatives $\partial z_{\theta}^{\rm DD}/\partial z_{\alpha}$ calculated by using expressions~(\ref{DDfunction}) are
\be
\la{PartialDerivativeDD}
z_{\alpha}\frac{\partial z_{\theta}^{\rm DD}}{\partial z_{\alpha}} = z_\alpha^{DD}\delta_{\alpha,\theta} 
 - \frac{4 \pi \beta e_\alpha e_\theta z_{\alpha} z_\theta}{\kappa_z^2}  
-\frac{(4 \pi \beta)^2 e_\alpha^2 e_\theta z_{\alpha} z_\theta}{\kappa_z^4} \sum_\gamma e_\gamma z_{\gamma}
\ee
where $\delta_{\alpha,\theta}$ is the Kronecker symbol.
For any set $\{ z_{\alpha} \} $ satisfying the pseudo-neutrality 
condition~(\ref{PseudoNeutrality}), these expressions become 
\be
\la{PartialDerivativeDDbis}
z_{\alpha}\frac{\partial z_{\theta}^{\rm DD}}{\partial z_{\alpha}} = z_\alpha \delta_{\alpha,\theta}  - \frac{4 \pi \beta e_\alpha e_\theta z_{\alpha} z_\theta}{\kappa_z^2}   \; ,
\ee
while the Debye-Dressed functions $z_\alpha^{\rm DD}$ reduce to $z_\alpha$. Inserting these results into Eq.~(\ref{DensityPressureADDbis}), 
we exactly recover the expressions~(\ref{DressingRecipe}) for the dressed densities.

\subsection{Comparison with other prescriptions ensuring electroneutrality}
\la{Sec:S42}

In order to compare the Debye-Dressing neutralization prescription~\eqref{DressingRecipe} with that of the Neutral Groups, it is useful to rewrite Eqn.~(\ref{DressingRecipe}) in a way similar to expressions~(\ref{DensityPressureANGter}), namely
\begin{align}
\la{DensityPressureADDter}
&\rho_i =  \sum_{\delta=1}^{\cS} D_{i,\delta} \; z_\delta \; \frac{\partial P_{\rm A}}{\partial z_{\delta}} (T;\{z_\gamma \}) 
 \quad \text{for} \quad i=1,...,\cS-1 \nonumber \\
&\rho_{\rm e}=   \sum_{j=1}^{\cS-1} Z_j \rho_j \; 
\end{align} 
with coefficients 
\be
\la{CoefficientsNGDensityB} 
D_{i,j}=
\begin{cases}
\displaystyle
 1-\frac{Z_i^2 z_i}{z_{\rm e} + \sum_{l=1}^{\cS-1}Z_l^2 z_l}    &\text{if  $j=i$} \\
 \displaystyle\rule{0mm}{8mm}
-\frac{Z_i Z_j z_i}{ z_{\rm e} + \sum_{l=1}^{\cS-1}Z_l^2 z_l}      &\text{for } j=1,...,\cS-1 \; , \; j \neq i  \\ 
\displaystyle\rule{0mm}{8mm}
 \frac{Z_i  z_i}{ z_{\rm e} + \sum_{l=1}^{\cS-1}Z_l^2 z_l}   &\text{if $j=\cS$ [i.e. $j={\rm e}$]}
\end{cases}. 
\ee
For two-component systems, like the hydrogen or the helium plasmas for instance, it turns out that 
$D_{1,1}=C_{1,1}=1/(Z+1)$ and $D_{1,{\rm e}}=C_{1,{\rm e}}=Z/(Z+1)$, so both recipes are equivalent. For 
systems with three or more components, like the hydrogen-helium mixture, these methods are no longer 
equivalent, at least for the choice~(\ref{RelevantActivity}) of the Neutral-Group activities. Nevertheless it is worthy to note that both 
prescriptions become equivalent if the approximate pressure $P_{\rm A}$ is consistent with local charge neutrality, i.e. 
if the undressed densities~(\ref{UndressedD}) do not carry a net charge $q_{\rm exc}=0$. {Of course, they become
exact for an exact expression of the pressure. }

\bigskip

Let us mention that yet another neutralization prescription has been used in the literature~\cite{Starostin2005}, which we call the Enforced-Neutrality prescription. Contrarily to the previous 
Neutral-Group or Debye-Dressed procedures, it does not rely on a general transformation valid for any approximate 
pressure $P_{\rm A}$. For a given set of nuclei activities $\{ z_i; i=1,...,\cS-1 \}$, it consists in choosing the 
electron activity $z_{\rm e}$ in such a way that the local charge neutrality for the densities directly calculated within 
the standard formulae is indeed observed. The particular value of $z_{\rm e}$ if found by solving a non-linear equation that is specific to considered model. Notice that this prescription disregards the pseudoneutrality condition~\eqref{PseudoNeutrality}.

\bigskip

Eventually, let us illustrate the various neutralization methods for a two-component system in the case of the following simple approximation for the pressure 
\be
\la{ApproximationPring}
\beta P_{\rm A}(\beta;z_1,z_{\rm e}) = z_1 + z_{\rm e} + \frac{\kappa_z^{3}}{12 \pi} 
\ee
with $\kappa_z=[4\pi \beta e^2 (Z^2 z_1 + z_{\rm e} )]^{1/2}$. The first two terms are nothing 
but the ideal Maxwell-Botzmann contributions, while the last 
term is the classical mean-field (or ring) contribution. The Neutral-Group and Debye-Dressed methods 
provide the same densities
\be
\la{DensityRingTCP}
\rho_{\rm n} =  z \left[ 1 + \beta e^2 Z \kappa_z/2   \right] \quad \text{and} \quad \rho_{\rm e}= Z \rho_{\rm n} 
\ee
where the subscript `n' refers to nuclei ($z_1 = z_{\rm n} = z, z_{\rm e}=Zz$).
These expressions also coincide with the exact small-activity expansion of $\rho_{\rm n}$ and $\rho_{\rm e}$ 
up to order $z^{3/2}$ included, which can be calculated within the diagrammatic series~(\ref{IX.QMG88}). Hence, the 
approximate EOS associated with~(\ref{ApproximationPring}) are identical in both methods, namely
$ P_{\rm A}^{\rm NG}(\beta;\rho_{\rm n},\rho_{\rm e})= P_{\rm A}^{\rm DD}(\beta;\rho_{\rm n},\rho_{\rm e})$.

\bigskip

Within the Enforced-Neutrality procedure, since
\be
\la{ApproximationPringDerivative}
z_1 \frac{\partial \beta P_{\rm A}}{\partial z_1} = z_1 + \beta e^2 Z^2 \kappa_z/2 \quad \text{and} 
\quad z_{\rm e} \frac{\partial \beta P_{\rm A}}{\partial z_{\rm e}} = z_{\rm e} + \beta e^2 \kappa_z/2 \; ,
\ee
if we set $z_1=z$, the electron activity $z_{\rm e}^{\rm EN}$ is such that 
\be
\la{ENactivity}
Z \left[z + \beta e^2 Z^2 \kappa_z/2 \right]  = z_{\rm e}^{\rm EN} + \beta e^2 \kappa_z/2  \quad \text{with} 
\quad \kappa_z=\left[4\pi\beta e^2(Z^2 z +z_{\rm e}^{\rm EN}) \right]^{1/2} \; ,
\ee
which can be recast as a cubic polynomial equation for $z_{\rm e}^{\rm EN}$. For $Z \neq 1$, 
$z_{\rm e}^{\rm EN}$ is different from the electron activity $Zz$ satisfying the pseudo-neutrality condition, 
and the resulting EOS $\beta P_{\rm A}^{\rm EN}(\beta;\rho_{\rm n},\rho_{\rm e})$ is different from the previous EOS
$P_{\rm A}^{\rm NG}(\beta;\rho_{\rm n},\rho_{\rm e})=P_{\rm A}^{\rm DD}(\beta;\rho_{\rm n},\rho_{\rm e})$. 
If $Z=1$, \textsl{i.e.} for the hydrogen plasma, $z_{\rm e}^{\rm EN}=z$ so the Enforced-Neutrality procedure is equivalent to the 
previous methods for the specific model~\eqref{ApproximationPring}. However, as soon as quantum corrections are added to the model, this equivalence no longer holds, even in the hydrogen plasma, because 
quantum effects involve particle masses $m_{\rm e}$ and $m_{\rm p}$ which are not identical.

\section{Derivation of approximate equations of state}
\la{Sec:S5}

The screened activity expansion for the pressure appears to be quite useful for 
constructing approximate expressions $P_{\rm A}(\beta;\{ z_i \})$ at moderate densities. 
In such regimes, recombination processes into chemical species made with three or more particles 
become important. The contributions of the relevant chemical species, including their interactions, are 
included in cluster functions. In a first step, the graphs which are expected to provide 
the main contributions are selected on the basis of physical arguments. In a second step, their 
contributions can be numerically computed by using simplified versions of $\phi$~\cite{Ballenegger2017}, while the 
functional integrations over loop shapes require the introduction 
of suitable quantum Monte Carlo techniques~\cite{Wendland2014}.

\subsection{Cluster functions associated with a given number of particles}
\la{ClusterFunctions}

The contributions of familiar chemical species can be easily identified in terms of specific diagrams 
in the screened activity expansion~(\ref{ScreenedMayerP}) of the pressure, following the method first
introduced for the particle densities~\cite{Alastuey2003}. It consists in rewriting the phase space measure of each loop $\cL$, as a sum over 
the number $q$ of elementary particles (nuclei or electrons) which are contained in $\cL$. Each graph ${\cal P}$ then generates an infinite number 
of graphs ${\cal P}[N_1,...,N_{\cal S}]$ with the same topological structure. Each $N_\alpha$ is the total number of 
particles of species $\alpha$, obtained by summing the particle numbers in all the loops of species $\alpha$. The 
corresponding loop phase-space integration becomes 
\be
\la{PhaseSpaceLoopFixedN}
\int D(\cL) \rightarrow \int \d \bx \int D_{q}(\bcX(\cdot)) 
\ee
Similarly to what occurs for the screened representation of particle 
densities~\cite{Alastuey2003}, ideal-like contributions of familiar chemical species $\cE[N_1,...,N_{\cal S}]$
made with $N_\alpha$ particles of species $\alpha$, $\alpha=1,...,{\cal S}$, are contained in the
sum of all the contributions of graphs ${\cal P}[N_1,...,N_{\cal S}]$.

\bigskip

Within the present formalism, the contributions of $\cE[N_1,...,N_{\cal S}]$ are dressed 
by the collective effects embedded in the screened potential $\phi$ as well as in the ring sum $I_{\rm R}$. 
The sum of the contributions of all 
graphs ${\cal P}[N_1,...,N_{\cal S}]$ for a given set $(N_1,...,N_{\cal S})$ defines a cluster function 
$Z[N_1,...,N_{\cal S}]$. It includes ideal-like contribution for the dressed chemical species $\cE[N_1,...,N_{\cal S}]$, 
as well as interactions between the chemical species resulting from the dissociation of $\cE[N_1,...,N_{\cal S}]$.   

\bigskip

Let us consider the case of the hydrogen-helium mixture ${\cal S}=3$, made with protons ($\alpha=1$), 
alpha-nuclei ($\alpha=2$) and electrons ($\alpha=3={\rm e}$). 
Hydrogen atoms are associated with graphs ${\cal P}[1,0,1]$ made with one proton and one electron, helium atoms with graphs ${\cal P}[0,1,2]$
made with one alpha-particle and two electrons, etc... 
For instance, $Z[0,1,2]$ accounts for a dressed atom ${\rm He}$ as well 
as interactions between (i) one ion ${\rm He}^+$ and one electron (ii) one alpha-nuclei and one electron. 
Also, $Z[2,0,2]$ describes a dressed molecule ${\rm H}_2$, interactions between two dressed atoms ${\rm H}$, etc...

\bigskip

In the zero-density limit, the cluster functions can be related to suitably defined bare partition 
functions of the chemical species in the vacuum~\cite{Alastuey2003}. We stress that the systematic prescriptions 
defining these cluster functions avoid double counting problems. Moreover, they properly account for the collective 
screening effects which ensure the finiteness of the bare partition functions, without introducing ad-hoc regularizations 
as in the phenomenological Planck-Larkin partition functions (see \textsl{e.g. }~\cite{Ebeling2017,Ballenegger2012}).
For instance, in the case of the hydrogen plasma made with protons ($\alpha=1$) and electrons ($\alpha=2$), the zero-density limit of $Z[1,1]$ gives rise to the bare partition function
$Z_{{\rm H}}$ of the hydrogen atom in the vacuum, which is close to the virial second-order function $Q$ first introduced 
by Ebeling~\cite{Ebeling1967}. Similar partition functions $Z_{{\rm H}_2^+}$, $Z_{{\rm H}^-}$ and $Z_{{\rm H}_{2}}$ for ions and molecules can 
be defined. They control the systematic corrections to Saha theory for a partially ionized atomic gas~\cite{Alastuey2008}.

\subsection{Simple scheme using the Debye-Dressing neutralization prescription}

In order to calculate the particle densities associated with a given $P_{\rm A}(\beta;\{ z_i \})$, 
use of the DD prescription is particularly attractive. Firstly, it is based on an important physical mechanism related to 
Debye screening. Secondly, the dressed densities~(\ref{DressingRecipe}) are given by 
a general expression which does not depend on the form of $P_{\rm A}(\beta;\{ z_i \})$. 
Eventually, the resulting EOS can be determined within the following scheme
which is simple to implement in practice. For fixing ideas, we illustrate this scheme for a three-component system
like the hydrogen-helium mixture for instance:

\begin{enumerate}

\item Consider various sets $(z_1,z_2,z_{\rm e})$ that satisfy the pseudo-neutrality condition~(\ref{PseudoNeutrality}), i.e. $z_{\rm e}=Z_1z_1 + Z_2z_2$. For each set, compute\\[-1.8em]
\begin{itemize}
\item the pressure $P_{\rm A}(\beta, z_1,z_2, z_{\rm e})$
\item the (DD) particle densities~(\ref{DressingRecipe}) through numerical partial differentiations of $P_{\rm A}$.
\end{itemize}

\item From the pressures and the associated densities computed at the previous step, determine the EOS $\beta P_{\rm A}^{\rm DD}(\beta;\rho_1,\rho_2)$.

\end{enumerate}

This scheme avoids having to invert the relation between the pseudo-neutral sets $(z_1,\, z_2,\, Z_1z_1 + Z_2z_2)$ and 
the nuclei densities $(\rho_1,\rho_2)$ for computing $\beta P_{\rm A}^{\rm DD}(\beta;\rho_1,\rho_2)$. Other approximate EOS would be obtained 
by using either the Neutral-Group method or the Enforced-Neutrality procedure. However, for approximate functions $P_{\rm A}(\beta;\{ z_i \})$ 
obtained within the diagrammatic series~(\ref{ScreenedMayerP}), the Debye-Dressing recipe is more directly related to a crucial mechanism at 
work than these methods. Hence, it can be reasonably expected to provide better EOS than the Neutral-Group or Enforced-Neutrality procedures.

\section{Conclusions et perspectives}
\label{sec:S6}

We have derived the screened activity series \eqref{ScreenedMayerP} of the pressure of a quantum multicomponent plasma, which provides a convenient route for computing the equation of state of such systems at low and moderate densities. We have demonstrated that this new series simplifies significantly the calculation of the EOS by reducing drastically the number of diagrams to be computed and by being more efficient for a numerical perspective since it avoids integrating term-by-term diagrams contributing to the particle densities. This representation is also quite promising for deriving approximate EOS for moderately dense plasmas. In particular it 
accounts, in a non-perturbative way, for the emergence of any chemical species, atoms, molecules, ions, which are formed through 
recombination processes of nuclei and electrons. 
Use of the screened activity expansion of the pressure offers a wide 
flexibility for various approximations, through the selection of relevant graphs associated with 
crucial mechanisms at work. Accurate approximations for the screening potential $\phi$, which simplify the task of computing such graphs, are also available~\cite{Ballenegger2017}. 

\bigskip
 
When devising an approximate theory, it is crucial to ensure that it is compatible with the local charge neutrality. We have devised two schemes for enforcing electroneutrality in approximate theories. The first scheme, the Neutral-Group (NG) neutralization prescription, is based on the Lieb-Lebowitz theorem which implies that the exact pressure depends on the activities only via neutral-group activities, which are variables with a clear physical interpretation. This prescription is very general and several implementations of this scheme are possible in plasmas with three or more components. It is straightforward to use since the corresponding densities are given by fully explicit formulae. The second neutralization scheme, the Debye-Dressing (DD) prescription, is also new and uses the Debye screening effect to enforce electroneutrality. More specifically, the appearance of a neutralizing polarization cloud around each particle is accounted for in that scheme at all orders in the particle activities at a mean-field classical (Debye-Hückel) level. The choice of a particular scheme is worthy of attention because it can affect the computed equation of state, as shown on a simple example. Contrary to the Enforced-Neutrality (EN) scheme which has been used previously, the NG and DD schemes do not break the pseudo-neutrality condition, which is often employed in EOS calculations in the grand-canonical ensemble. The latter two schemes being fully explicit, they do not require solving any equation specific to the studied system. The DD prescription is closely related to the NG scheme. 
Whether the DD prescription is a special case of a NG prescription for a specific choice of basis for the neutral groups remains on open question. When calculating an approximate equation of state by using the new diagrammatical series~\eqref{ScreenedMayerP} for the pressure, the DD prescription should be preferred because it is based on a physical phenomenon and because double-counting of screening effects can be avoided by a proper selection of the retained diagrams in the pressure series.

\bigskip

Eventually, the methods presented in this paper will be applied to derive accurate approximate equations of state for hydrogen and hydrogen-helium mixtures at moderate densities. The EOS of such plasmas can be studied by computing the screened Mayer diagrams using analytical and numerical techniques.
The cluster functions for fixed number of particles, defined by summing Mayer diagrams with a constraint on the total number of particles, play a central role in such calculations~\cite{Ballenegger2017,Alastuey2012a,Alastuey2008}. A partial account of calculations along the Sun adiabat is given in Refs.~\cite{WendlandThesis,Ballenegger2018}. A more systematic study including denser regimes will be published elsewhere.

\section*{Acknowledgements}

Financial support from the CNRS (contract 081912) and from the Conseil r\'egional de Franche-Comt\'e (contract 362887) are gratefully acknowledged.

\end{document}